# A tutorial review: probing molecular structure and dynamics with CEI and LIED

**Kasra Amini**[1]

[1]Max-Born-Institut, Max-Born-Straβe 2A, 12489 Berlin, Germany

E-mail: kasra.amini@mbi-berlin.de

## Abstract

Knowledge of the molecular structure is key to understanding the function of participating molecules in photo-induced chemical reactions. Visualizing the nuclear dynamics of a photochemical reaction requires an ultrafast measurement technique that can identify the location of atoms in molecules with atomic Ångstrom spatial resolution evolving on the nuclear (i.e. hundreds of femtosecond) timescale. Coulomb explosion imaging (CEI) and laser-induced electron diffraction (LIED) offer the required sub-Ångstrom spatial and tens of femtosecond temporal resolution to track in real-time changes in the molecular structure. In this tutorial review, details of the tools, analysis procedures, exemplary previous results and future perspectives of both CEI and LIED techniques are described.

## 1. Introduction

Photo-induced chemical reactions play a crucial role in many important processes in nature such as vision [1,2] and photosynthesis [2]. Predicting the outcome of such photochemical reactions requires knowledge of both the nuclear and electronic structure of molecules. A key objective of ultrafast chemical physics is to measure these dynamics in real-time as they occur during a chemical photochemical reaction. Various experimental observables provide a means to measure nuclear dynamics (scattering of electrons [3–46] or X-ray photons [15,16,47–60], correlating the emission directions of charged particles after photoionization [61–77,77–102]) and electronic dynamics (absorption of light [103–106] and the emission of electrons with specific excess energies after photoionization[106–108]) in real-time. The focus of this tutorial review is on the nuclear dynamics of photochemical reactions studied by two recently developed ultrafast methods: *Coulomb explosion imaging (CEI)* [61–102] and *laser-induced electron diffraction (LIED)* [3–8,10,109–127].

One of the fundamental principles of photochemistry is the concept of *multi-dimensional potential energy surfaces (PESs)*. This stems from the Born-Oppenheimer approximation in quantum chemistry, where the energies of each electronic state are calculated at various specific molecular geometries. Chemical reactions evolve on this PES by following the energetic minimum of the PES. Fig. 1 shows the potential energy curves (PECs; two-dimensional cuts of a PES) of the electronic states participating in the photodissociation of a generic molecule AB. Here, the AB molecule is photoexcited by an ultraviolet pulse (blue arrow) from its ground electronic state ($S_0$, black curve) to its first excited state ($S_1$, red curve) to generate a nuclear wave packet (NWP, see blue gaussian). The NWP reaches the minimum of the $S_1$ state but it possesses sufficient energy to reach large A-B internuclear distances, $R_{AB}$, and the subsequent photodissociation of the A-B bond.

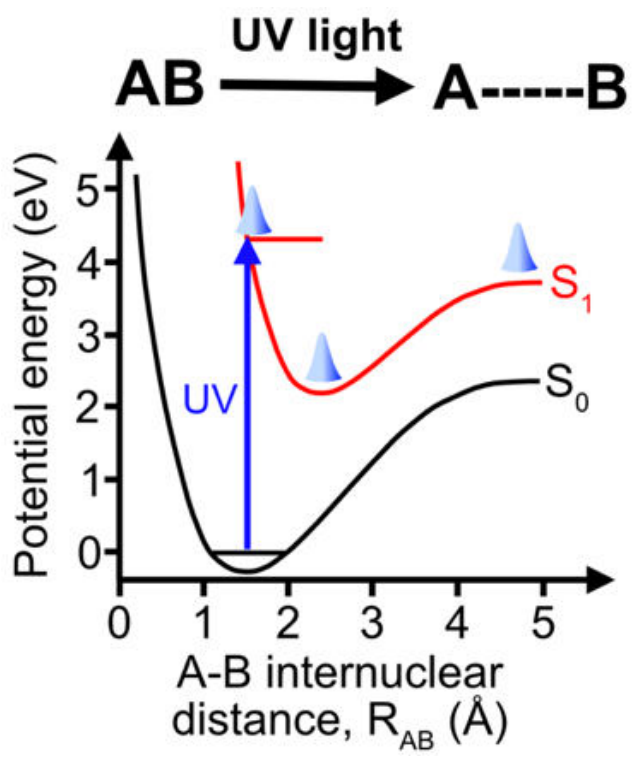

**Figure 1.** Potential energy curves (PECs) of the ground ($S_0$, black) and first excited ($S_1$, red) electronic states involved in the photodissociation of a generic molecule AB following its excitation by an ultraviolet (UV, blue) pulse to generate a nuclear wave packet (blue gaussian).

Experimentally, the nuclear and electronic dynamics of gas-phase molecules is studied using the pump-probe technique. [128–130] Here, an optical "pump" pulse initiates the photochemical reaction of interest, providing the start time of the reaction, $t_0$. A subsequent "probe" pulse measures a signature of the molecular and electronic structure at various times $t$ after the pump pulse, called the pump-probe delay, $\Delta t = t_0 - t$. A series of snapshots is recorded at a variety of different pump-probe delays to generate a "molecular movie" of the ensuing photochemical reaction. Nuclear dynamics in gas-phase photochemical reactions occur on the hundreds of femtosecond timescale (fs, 1 fs = $10^{-15}$ s) with changes in the molecular structure occurring on the tens-to-hundreds of picometre (pm, 1 pm = $10^{-12}$ m) sub-atomic spatial scale. There are two approaches to study nuclear dynamics: (i) spectroscopy and (ii) imaging techniques.

In spectroscopy, energy changes between the initial and final state are measured, and the molecular structure can be indirectly retrieved by comparison to calculated spectra corresponding to specific molecular structure configuration(s). For example, microwave [131–135] and infrared [132,136] spectroscopies can determine the molecular structure with very high precision (<1-pm). However, these spectroscopic methods rely on narrowband lasers and so their time-resolved analogues on the timescale of nuclear dynamics (i.e. hundreds of femtoseconds) are not readily available. Ultrafast spectroscopic methods based on broadband pulses such as transient absorption [103–106] and time-resolved photoelectron [106–108] spectroscopies have reported information on the nuclear dynamics of photochemical reactions. These ultrafast analogous of spectroscopy measurements also provide indirect information on the nuclear structure due to its reliance on comparison to calculated spectra.

In imaging, the position of atoms in a molecule can be directly visualized and therefore the molecular structure can be directly retrieved. In the last few decades, several ultrafast imaging techniques have been developed, driven by broadband laser pulses, that achieve both the spatial (<1-Å) and temporal (hundreds of fs) resolution required to directly visualize changes in the positions of atoms within molecules undergoing a photochemical reaction. The most commonly used ultrafast imaging techniques are ultrafast electron diffraction (UED) [4–15,3,16,17,20,18–45] and ultrafast X-ray diffraction (UXD) [15,16,47–60]. In UED and UXD, the electrons and X-ray photons scatter against the inner-shell core-lying electrons close to the nucleus of atoms in molecules. The scattered electrons or X-rays are then detected on a position-sensitive detector, with structural information embedded in their momentum distribution that is measured in reciprocal space (i.e. where the scattering measurement is performed). The Fourier transform of the measured sinusoidally oscillating molecular interference signal generates the radial distribution in real space (i.e. where the target structure is visualized), providing information on the bond lengths and internuclear distances of atoms in the molecule. The recent development of the latest generation X-ray free-electron lasers (XFELs) at large-scale institute facilities provide significantly higher photon flux (e.g. $10^{14}$ photons/s) [15] than current tabletop X-ray sources (e.g. $10^8$ photons/s) [137], enabling XFEL-based X-ray diffraction to match or even surpass electron diffraction in terms of the achievable quality of detected diffraction signal [15]. In the framework of easily accessible tabletop sources, electron diffraction is preferred over X-ray diffraction because of the orders-of-magnitude higher scattering cross-section and the significantly smaller de Broglie wavelength achievable than its photon counterpart.

The structure and dynamics of gas-phase molecules and condensed matter materials (e.g. solid-state and liquid samples) can be imaged using a UED set-up. In this set-up, an ultraviolet laser pulse impinges on a metal cathode surface (e.g. copper or gold), generating electron pulses due to the photoelectric effect. The electrons are then accelerated to the keV or MeV kinetic energy range over a short distance. The spatial and temporal characteristics of the multi-electron pulse is manipulated by magnetic and radiofrequency (RF) fields, respectively. The temporal focus is located at the sample position whilst the spatial focus is usually located on the detector to ensure the optimal reciprocal space (and thus real space) resolution at detection. The Coulombic repulsion between the many electrons in the electron pulse, called the space-charge effect, leads to a significant broadening in the electron pulse's temporal and energy distribution. In gas-phase UED studies, the space-charge broadening is compensated for by applying a linear velocity chirp to temporally compress the electron pulse using the linear portion of external compression fields (e.g. radiofrequency or terahertz optical fields).[25,33,35,40,42,43] Time-resolved gas-phase UED measurements have been reported with a temporal resolution of <250 fs in keV non-relativistic [25,40,44,45] and <150-fs in MeV relativistic [11,13,12,14] UED. To investigate photo-induced chemical reactions occurring on shorter timescales (e.g. the ring-opening of thiophenone in <500-fs [138]), further improvements in the temporal resolution of current UED techniques is required. Although an in-depth review of UED and UXD methods is beyond the scope of this tutorial review, further details can be found in Refs. [3,9,12,27,46] and [50,59], respectively, as well as a comparison between the two methods in Ref. [15].

The more recently developed techniques of CEI [61–102] and LIED [3–8,10,109–127] are the main focus of this tutorial review. Both CEI and LIED offer several advantages over UED and UXD, such as: (i) better temporal resolution, (ii) they are table-top set-ups that are more compact and cost much less than the facility-based UED and UXD set-ups. The purpose of this tutorial review article is to give a brief overview of the tools and analysis steps required to perform laser-induced CEI and LIED in retrieving the geometric structure and molecular dynamics of gas-phase molecules in section 2. Details and previous relevant work of CEI and LIED are given in sections 3 and 4, respectively. A general summary and future outlook of these and other methods will be briefly given in section 5.

## 2. Tools required to perform CEI and LIED

CEI and LIED rely on a number of essential components, including a cold molecular beam target, the generation and propagation of intense femtosecond laser pulses, the strong field ionization process, and fast charged particle detection schemes. In this section, we will provide a short overview of the working principle of the instruments and tools at the heart of these two techniques. We note however that the aim of this section is not to provide an exhaustive discussion of each of these elements separately but instead to provide the reader with the necessary knowledge regarding the physical principles of the tools typically used in LIED and CEI experiments. We will also introduce the strong-field ionization processes, which plays an important role in both techniques.

### 2.1 Strong-field ionization

Both CEI and LIED rely on the strong-field ionization of the molecular ensemble by intense femtosecond laser pulses. Fig. 2 shows the effect of an intense femtosecond laser pulse on an atom's Coulomb potential. In the field-free case (Fig. 2a), a valence electron (green circle) is bound by the Coulomb potential (red dashed) with a binding energy equal to its ionization potential, $I_{\mathrm{p}}$. As the field strength of a low frequency laser pulse increases to an appreciable level, the Coulomb potential becomes perturbed and ionization by a low-frequency laser field becomes possible. In the case of a small perturbation, the ionization potential is reduced such that ionization by the absorption of multiple photons can occur, referred to as the multiphoton ionization regime (Fig. 2b). Once the field strength reaches $10^{13}$ Wcm$^{-2}$ or higher, the Coulomb barrier is significantly distorted such that the bound electron experiences a small Coulomb barrier through which the electron can tunnel through. This ionization regime is referred to as the tunnel ionization regime (Fig. 2c). At even higher intensities, the Coulomb barrier is supressed leading to over-the-barrier ionization of the atom (Fig. 2d). The ionization regime of operation can be identified by the Keldysh parameter, $\gamma$, given by[139,140]

$$\gamma = \sqrt{\frac{I_{\mathrm{p}}}{2U_{\mathrm{p}}}}, \tag{1}$$

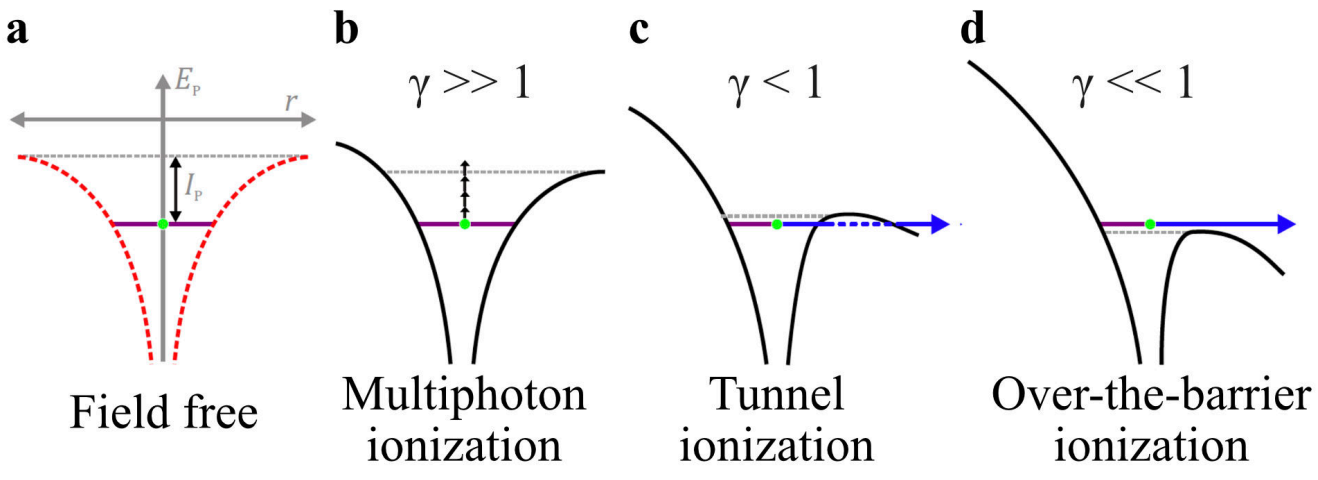

**Figure 2.** Ionization of an atom by a strong laser field. (a-d) The atomic Coulomb potential for the field-free (a), multi-photon (b), tunnelling (c), and over-the-barrier ionization (d) regimes. The binding energy, $I_{\mathrm{p}}$, Keldysh parameter, $\gamma$, photon energy, $E_{\mathrm{p}}$, and reaction coordinate, $r$, are indicated. Figure adapted from [69].

which depends on the ionization potential of the atom or molecule, $I_{\mathrm{p}}$, and the ponderomotive energy, $U_{\mathrm{p}}$, of the laser field (i.e. the average kinetic energy of a free electron in an oscillating electric field of the laser pulse) defined as [141]

$$U_p = \frac{Ie^2\lambda^2}{8\pi^2 m_e \epsilon_0 c^3} \propto I\lambda^2 , \tag{2}$$

where $I$ is the electric field peak intensity, $\lambda$ is the central wavelength of the laser pulse, $m_{\mathrm{e}}$ is the mass of an electron, $\epsilon_0$ is the permittivity of vacuum, and $c$ is the speed of light.

### 2.2 Single-photon ionization versus strong-field ionization

Single-photon ionization corresponds to the ionization of an atom or a molecule by the absorption of a single photon. In most atoms and molecules, single photon ionization takes place when the photon energy is higher than the ionization potential (or the binding energy) of the electron in the atom or the molecule. The removal of a valence electron from an atom or a molecule (i.e. the highest occupied electronic orbital) requires photons typically in the extreme ultraviolet spectral range. At higher photon energy (e.g. soft and hard X-rays), an electron can be removed from a deeper electronic orbital (core-shell ionization) and can be accompanied by an Auger decay. In this decay, an electron from a higher energy level refills the inner-shell vacancy and the excess of energy is transferred to a second electron that is emitted. The emission of the second electron is referred as Auger electron. Thus, the absorption of a single high-energy photon can result in the removal of several electrons. At sufficiently high intensity ($\gg 10^{13}$ Wcm$^{-2}$), multiphoton absorption can also take place, leading to the emission of a large number of electrons either by sequential single-photon absorption process or by simultaneous absorption of two or more photons. This intensity regime can be easily achieved at free-electron lasers (FELs) [142–150] which provide high-flux high-energy photon sources that site-selectively ionize the molecule to very high charge states. [76,77,82,88,92] FELs provide high-flux high-energy photons that can site-selectively ionize molecules to very high charge states by sequential or direct multiphoton ionization process from core shell orbitals. For example, at a photon energy of 107 eV, an iodine-containing molecule such as $CH_3I$ can be site-selectively ionized on the iodine atom since the absorption cross-section is significantly higher for the iodine atom (3.57 Mb) than the rest of the molecule ($CH_3$; 0.53 Mb). The high flux offered by FELs enables the absorption of multiple high-energy photons at the iodine atom to achieve an initial high charge state initially localized at the iodine atom of the molecule. In general, to perform strong-field experiments, a target delivery system, femtosecond laser pulses and charged particle detection schemes are required.

### *2.3 Molecular beams*

Molecular beam gas jets enable the measurement of gas-phase molecules with intense, ultrafast laser pulses without the common thermal heating issues experienced in the study of solid-state samples with such intense pulses since the gas sample is continuously renewed. The typical recoil momenta of ions and electrons generated by strong-field ionization ranges from << 1 a.u. to tens of a.u.. However, the momentum spread of atoms in an atomic gas jet at room temperature and pressure is on the same order as typical recoil momenta of ions and electrons (i.e. around 10 a.u.). Thus, a method is required to cool the gaseous molecules such that the recoil momenta can be well-resolved. This can be achieved with a molecular beam gas jets that is generated by the adiabatic isentropic expansion of gas-phase molecules from a container maintained at high pressure ($p_1 \geq 1$ bar) through a small orifice, called a nozzle, into an evacuated chamber of low pressure ($p_2 \leq 10^{-5}$ mbar), as shown in Fig. 3.[151–153]. A supersonic gas jet is formed in the evacuated chamber where the internal energy (i.e. rotational, vibrational energy) of a non-directional thermal gas source is transferred into translational energy to generate a highly directional gas jet with a higher velocity, $v_\mathrm{j}$. For example, the central velocity of ammonia molecules is increased from ~500 m/s to ~900 m/s after supersonic expansion of a thermal gas reservoir ($p_1 \sim 1$ bar) into an evacuated chamber ($p_2 \sim 10^{-5}$ mbar) and its subsequent collimation with a skimmer, as shown in Fig. 3.[153] The velocity of the directed gas is greater than the speed of sound, $v_\mathrm{s}$, and its velocity is characterized by the Mach number[154]

$$M_\mathrm{a} = \frac{v_\mathrm{j}}{v_\mathrm{s}} = \sqrt{\frac{2}{\gamma - 1}\left(\frac{T_1}{T_2} - 1\right)}, \qquad (3)$$

where $T_1$ is the initial temperature of the gas in the reservoir bottle, $T_2$ is the temperature of the supersonically expanded gas jet, and $\gamma$ is the heat capacity ratio $\gamma = c_p/c_V$ of the heat capacity at constant pressure and constant volume, $c_p$ and $c_V$, respectively. The adiabatic, isentropic expansion is described by [152]

$$\frac{T_1}{T_2} = \left(\frac{p_1}{p_2}\right)^{\gamma-1/\gamma}, \qquad (4)$$

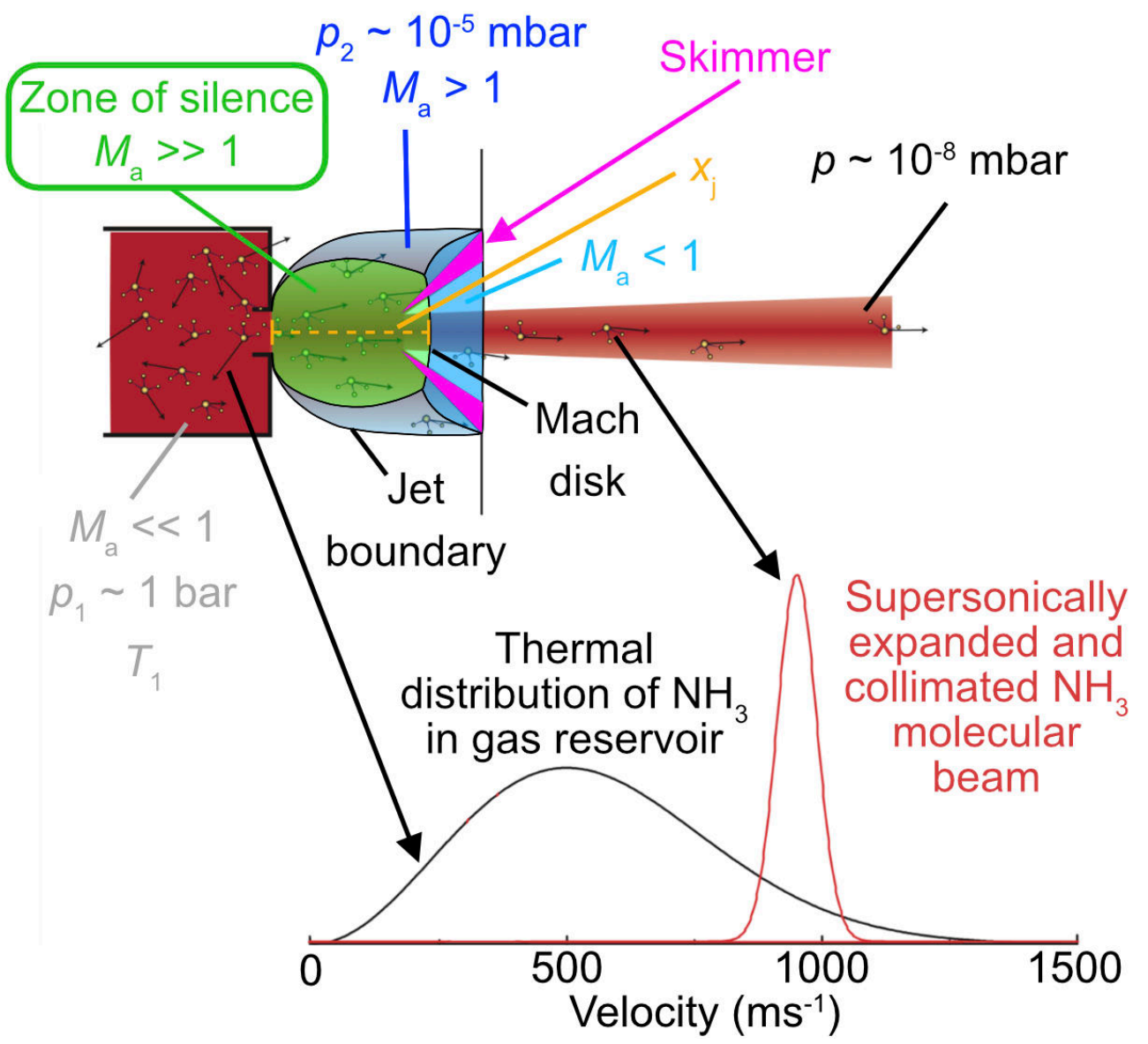

**Figure 3.** Supersonically expanded molecular beams. Illustration of ammonia ($NH_3$) molecules in a gas reservoir of a nozzle ($p_1 \sim 1$ bar) that are supersonically expanded ($p_2 \sim 10^{-5}$ mbar) and collimated ($p \sim 10^{-8}$ mbar) in evacuated chambers. The translational velocity distribution of ammonia molecules before (black) and after (red) supersonic expansion and collimation are shown. Arrows indicate the centre-of-mass travel direction. The Mach, $M_\mathrm{a}$, cone formed after the expansion of a free gas jet (with initial pressure $p_1$, temperature $T_1$ and $M_\mathrm{a} \ll 1$) into an evacuated chamber. The largest $M_\mathrm{a}$, and therefore coldest jet, is formed in the zone of silence (see green shaded). The distance of the zone of silence boundary from the nozzle position is given by $x_\mathrm{j}$ (see yellow). Figure adapted from [153] and [155].

where a significant decrease in pressure after a supersonic expansion (i.e. $p_1 >> p_2$) leads to a lower gas jet temperature ($T_2$) given by Eqn. (4), generating a significantly fast gas jet with $M_\mathrm{a} >> 1$ according to Eqn. (3). Decreasing the temperature of the gas subsequently corresponds to lowering the internal energy of molecules, $U_2$, as described by [151,152]

$$U_2 = \left(\frac{3}{2}\right) k_\mathrm{B} T_2, \qquad (5)$$

where $k_\mathrm{B}$ is the Boltzmann constant. The fastest, and therefore coldest, gas jet is generated at a distance $x_\mathrm{j}$ away from the nozzle of diameter $d$ as given by [154]

$$x_\mathrm{j} = d * 0.67 \left(\frac{T_1}{T_2}\right)^{\frac{\gamma}{2(\gamma-1)}} \qquad (6)$$

and is called the zone of silence (green shaded region, Fig. 3). A skimmer (pink, Fig. 3) can be placed within the zone of silence to avoid the supersonic gas jet from collapsing and to collimate the molecular beam. Mixed molecular beams consisting of a target molecule seeded with a lighter "carrier" gas (e.g. noble gases such as helium) can lead to the acceleration of the target molecule, generating a faster jet velocity and thus a colder molecular beam.

### *2.4 Generation of intense, femtosecond laser pulses*

The investigation of ultrafast photo-induced chemical reactions requires the use of ultrashort laser pulses with a pulse duration that is much shorter than the dynamics under investigation. Producing ultrashort laser pulses requires the generation of light with a broad range of frequency components, $\omega$, as dictated by the time-bandwidth product for a Gaussian pulse as given by

$$\Delta\omega\Delta\tau \geq 0.441, \quad (7)$$

where $\Delta\omega$ is the spectral width and $\Delta\tau$ is the pulse duration. Fig. 4a shows an optical cavity that includes a generating medium (typically a crystal) between two mirrors that generates a laser pulse with a broad range of different frequency components, $\omega_n$, as illustrated by the frequency spectrum in Fig. 4b. Here, standing waves of different frequencies resonate around the cavity composed by two mirrors separated by the so-called resonant length, $L_R$, with their constructive interference eventually generating coherent pulses in narrow temporal windows. Increasing the number of interfering standing waves further decreases the time window in which constructive interference takes place, leading to the further reduction of the pulse duration. The pulse duration and the spectral width or bandwidth of the emitted light is characterized by the FWHM of the temporal and frequency spectrum, as indicated in Fig. 4b. The lowest theoretical pulse duration that a bandwidth $\Delta\omega$ could support is given by $\Delta\tau = 0.441/\Delta\omega$, and is called a transform-limited pulse, which follows the Fourier transform relationship between frequency and time. To achieve the lowest pulse duration all frequency components must be temporally overlapped, which implies that there is a fixed phase relationship between the frequency components. The lowest pulse duration for a fixed phase relationship will be obtained when the phase is constant across the spectrum of the pulse, leading to a so-called transform-limited pulse. To further emphasize the importance of a fixed phase relationship, the coherent sum of ten frequency components or "modes" is shown in Fig. 4c. Each mode oscillates sinusoidally with an electric field intensity, $E_n(t)$, that can be described by the following expression

$$E_n(t) = E_0 \sin\big(\omega_n t + \phi(\omega_n)\big), \quad (8)$$

where $\omega_n$ is the central frequency of the mode, and $\phi(\omega_n)$ is the phase of the standing wave which defines the starting position of the wave in frequency space. Each mode can be thought of as behaving like an "independent laser", each emitting light at slightly different frequencies. In reality, all modes can constructively and destructively interfere with each other. The coherent sum of all modes with a fixed phase relationship leads to the constructive interference of all modes at specific points in time (see black vertical dotted line in Fig. 4c), generating a pulse train containing intense, ultrashort pulses (see blue sum distribution in Fig. 4c). Modes with a random phase relationship generate an overall pulse with a random electric field structure (see green sum distribution in Fig. 4c). The concept of a fixed phase relationship between modes is the fundamental principle of mode-locking[156] and ultrafast laser physics. In general, a broader frequency spectrum with a larger number of frequency components will generate a coherent signal that spans a narrower time range with a higher peak intensity.

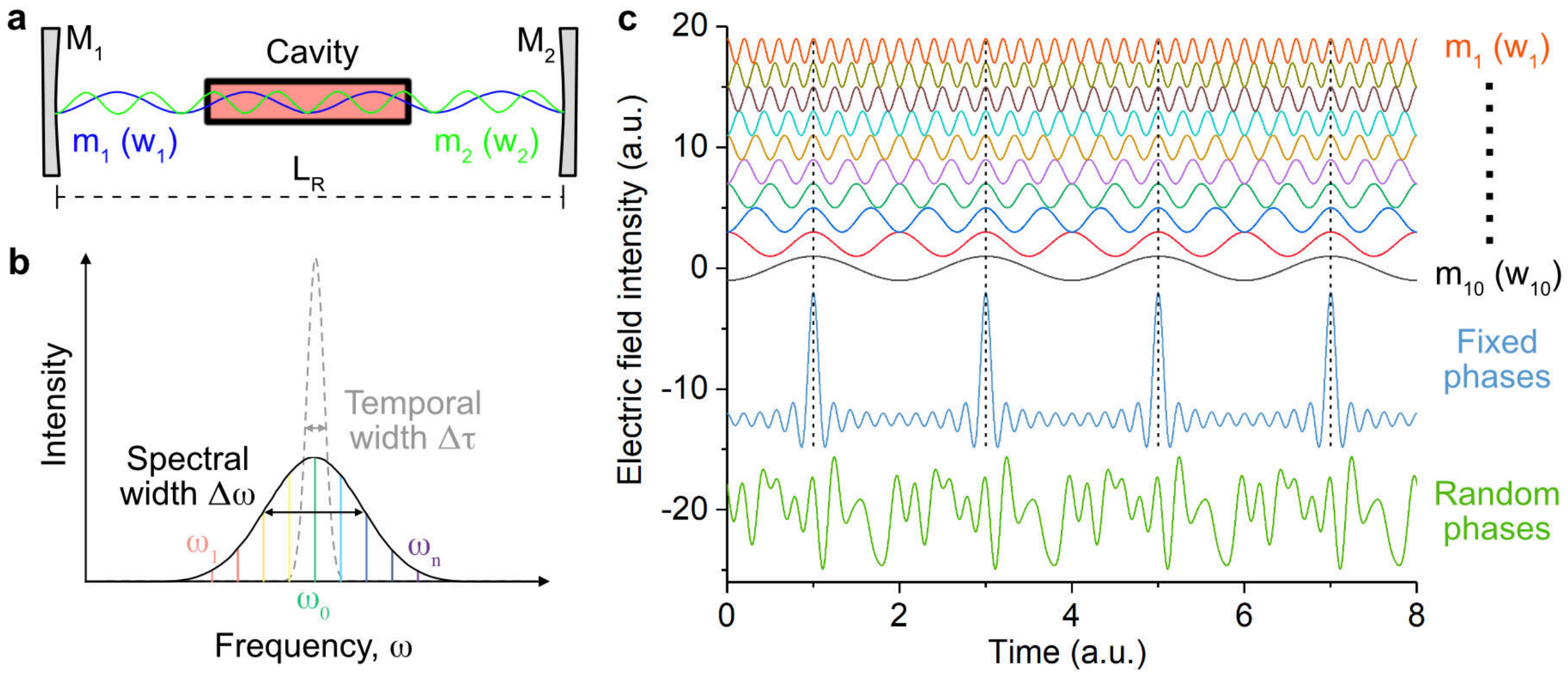

**Figure 4.** Generation of intense, femtosecond laser pulses. **(a)** A laser pulse generated using an optical cavity (orange) and two mirrors, $M_1$ and $M_2$. A standing wave is generated between two mirrors separate by a distance $L_R$. The wave is generated with a frequency $\omega_n = (n \cdot c)/(2 \cdot OPL)$ where $n$ is the mode number, $c$ is the speed of light, and $OPL$ is the optical path length given by $OPL = \eta * L_R$ where $\eta$ is the refractive index. **(b)** The frequency spectrum of a broadband ultrafast laser pulse. Individual frequency components, $\omega_n$, for $n$ components are indicated as vertical-coloured lines. The corresponding temporal distribution of the pulse is shown as a grey-dashed distribution. The spectral and temporal widths of the pulse, $\Delta\omega$ and $\Delta\tau$, respectively, are indicated and are the full width at half maximum (FWHM) values. **(c)** Ten modes, $m$, of different frequencies, $m_1(\omega_1)$ to $m_{10}(\omega_{10})$, as a function of time. The sum of all modes with a fixed phase (blue) and random phase (green) relationship is shown.

So far, the generation of femtosecond laser pulses using mode-locked oscillators have been discussed, which often generate relatively low pulse energies (i.e. nanojoules or lower). The pulse generated by the oscillator can be amplified by a process called chirped pulse amplification (CPA). In this process, the femtosecond pulse is first chirped and temporally stretched to the picosecond or nanosecond timescales using a strongly dispersive medium such as a grating pair. The stretched pulse is then amplified by passing through an amplifying medium to achieve a significantly larger pulse energy (i.e. millijoules or higher). The longer pulse duration reduces the peak power to avoid nonlinear pulse distortion or damage to the amplifying medium. The amplified, stretched pulse is then compressed to its initial pulse duration using a second grating pair with an opposite dispersion to that of the first grating pair. The signal of the femtosecond laser pulse can also be amplified using optical parametric amplifiers (OPAs). Here, the signal pulse of frequency $\omega_2$ propagates through a nonlinear crystal with a pump pulse of higher frequency $\omega_3$. After propagation, the pump photons are converted into lower-energy signal photons accompanied by the same number of idler photons of frequency $\omega_1$. This is due to energy conservation (i.e. $\omega_3 = \omega_1 + \omega_2$) which also ensures that the crystal material is not heated during the parametric amplification process. The efficiency of the parametric amplification process is further improved by using stretched pulses (as in the CPA concept) before amplification by an OPA through the optical parametric chirped-pulse amplification (OPCPA) process [157,158]. In this OCPPA process, the stretched pulses enable many more photons to be used during the parametric amplification process with a nonlinear crystal (as used in an OPA) which replaces the amplifying medium in the CPA process.

The wavelength of the amplified femtosecond pulses can also be modified to shorter (ultraviolet) or longer (mid-infrared) wavelengths using the OPA process by frequency doubling and tripling or by sum and difference frequency generation. For example, a mid-infrared (3100-nm) pulse can be generated through difference frequency generation by frequency mixing the two outputs (1030-nm and 1550-nm) from an erbium-based fiber laser [157,158]. Whereas an ultraviolet (267-nm) pulse can be generated through sum frequency generation by frequency mixing the fundamental (800-nm) and second harmonic (400-nm) of an 800-nm Ti:Sapphire laser system.

### *2.5 Generation of XUV and X-ray radiation: free-electron lasers*

Traditional lasers work very well in the optical spectral range since most amplifying media have a transition in this spectral range. For many applications, the use of extreme ultraviolet (XUV) and X-ray pulses is generally advantageous, for instance for femtosecond X-ray crystallography [159]. Ultrashort pulses of short wavelength radiation can be generated using (i) free electron laser (FEL) [142–150] technology that relies on the self-amplified spontaneous emission (SASE) process or (i) highly non-linear processes such as high-harmonic generation (HHG)[105,109,121,160–162]. Although details of the HHG process is beyond the scope of this tutorial review, further information can be found in Refs. [105,109,121,160–162].

The operating principle of FELs is shown in Fig. 5. Here, an ultrafast laser pulse irradiates a photocathode to generate an intense electron beam with a short pulse duration which are then accelerated to relativistic kinetic energies using linear accelerators [142,144]. The relativistic electron beam is injected into a long, periodic arrangement of dipole magnets with alternating polarity, called an undulator, that causes the electron trajectories to oscillate and produce spontaneous emission of radiation. At the beginning of the undulator, the electrons enter with random phase and thus the emitted radiation is incoherent, like synchrotron radiation. As the electron beam traverses through the undulator, it interacts with its own emitted radiation which leads to a so-called microbunching of the electron bunch. This has the effect of grouping electrons together with a separation approximately equal to the wavelength of the emitted radiation. After propagation in the undulator, the electron beam is separated into individual bunches that are separated by a distance equivalent to the wavelength of the emitted radiation. The large increase of the density of electrons in the individual bunches over a narrow time window defined by the wavelength of the emitted light leads to the amplification of the emitted synchrotron radiation. FELs are tunable in photon energy over a large spectral range, from the terahertz to the hard X-ray domain. However, since FELS rely on the amplification of synchrotron radiation, which is an incoherent process, significant shot-to-shot variations in the pulse energy, arrival time, $x$- and $y$-pointing directions and mean photon energy of the FEL pulse exist. This is somewhat improved using beam monitoring and correction systems [81] as well as seeding FELs with an external optical laser such as the FERMI FEL [163].

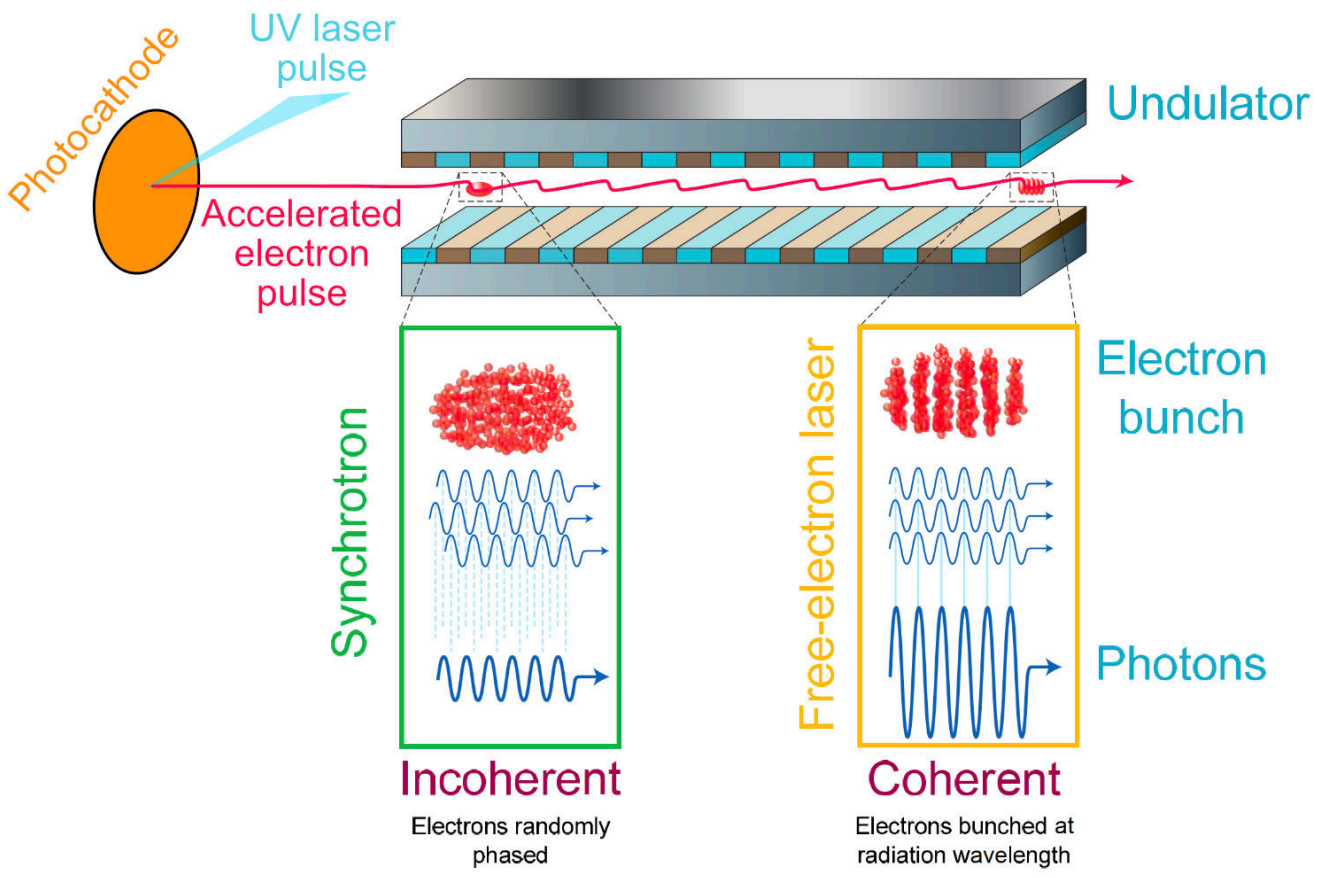

**Figure 5.** Operating principle of a free electron laser. An ultraviolet optical laser pulse impinges a photocathode to generate a pulsed electron beam that is accelerated to the GeV kinetic energy range. The relativistic electron beam is injected into an undulator that is composed of dipole magnets with alternating polarity causing the electron beam to oscillate, leading to the spontaneous emission of radiation. The initially incoherent electron beam is microbunched into coherent electron bunches as it travels across the undulator. The microbunching occurs by the interaction with its own emitted radiation, with the electron bunches separated by a distance determined by the wavelength of the emitted radiation. Figure adapted from [144].

## *2.6 Propagation of intense, femtosecond laser pulses*

In order to use ultrashort laser pulses, the laser beam needs to be propagated and steered to an experimental end station, which is typically achieved using a number of optical elements (e.g. mirrors, lenses, beam splitters etc) which can affect the laser pulse. In addition, the simple propagation of the ultrashort laser pulse in air (and in any material in general) can already alter the properties of the laser pulse due to the so-called dispersion. The dispersion of a medium describes how fast light travels in the material and can be expressed by

$$v_0(\omega) = \frac{c}{n(\omega)}, \tag{9}$$

where $v_0(\omega)$ is the phase velocity of the wave at frequency $\omega$ in the material with a refractive index $n(\omega)$, and $c$ is the speed of light. From this expression, waves with different frequencies propagate at different speed due to the dependence of the refractive index on the frequency (see Fig. 6a). When the refractive index decreases with increasing wavelength (the most common case), the medium is in the normal dispersion regime. In contrast, a medium has an anomalous dispersion when its refractive index increases with increasing wavelength. For a material with a normal dispersion, high frequency components will propagate at a slower speed through the material as compared to low frequency components. The propagation of an ultrashort pulse in a dispersive medium will therefore cause a frequency-dependent phase shift of the different frequency components, eventually leading to the temporal broadening of the pulse, which deteriorates the time resolution that can be achieved in a time-resolved measurement.

To describe the effect of dispersion from the propagation of an ultrashort laser pulse inside a medium, the electric field associated with the pulse can be written as a product of a pulse envelope and a carrier wave given by

$$E(\omega) = E_0\, g(\omega)\, cos(\omega_0 t + \phi(\omega)), \tag{10}$$

where $E_0$ is the maximum amplitude of the electric field, $\omega_0$ is the central frequency, and $\phi(\omega)$ is the carrier envelop phase that can be expressed in a Taylor series expansion around the central frequency $\omega_0$ as

$$\begin{aligned}\phi(\omega) = \phi_0 &+ \left\{\frac{1}{1!}\frac{\partial\phi}{\partial\omega}\bigg|_{\omega_c}(\omega-\omega_c)\right\} \\ &+ \left\{\frac{1}{2!}\frac{\partial^2\phi}{\partial\omega^2}\bigg|_{\omega_c}(\omega-\omega_c)^2\right\} \\ &+ \left\{\frac{1}{3!}\frac{\partial^3\phi}{\partial\omega^3}\bigg|_{\omega_c}(\omega-\omega_c)^3\right\} \\ &+ \cdots\end{aligned} \tag{11}$$

where $\phi_0$ is the absolute phase which determines the position of the electric field maximum relative to a time reference.

The second term contains the group delay, $T_g = \frac{\partial\phi}{\partial\omega}$, and describes the linear dependence of the phase $\phi$ with the frequency $\omega$. This term gives the amount by which the envelope is delayed in time without the carrier or envelope shape being modified. For example, a negative $T_g$ corresponds to a decrease in the arrival time of the pulse envelope at all frequencies (see green in Fig. 6d) because the absolute phase decreases with increasing frequency (see green in Fig. 6c).

The third term contains the group delay dispersion, GDD = $\frac{\partial^2\phi}{\partial\omega^2}$ , often called the second order dispersion as it describes the quadratic dependence of the phase $\phi$ with frequency $\omega$. To understand the concept of GDD (and so-called "chirp"), it is helpful to consider two cases where the GDD is either positive or negative. For positive GDD, the high frequency components are delayed in time such that the low frequency components arrive first at the material (see blue in Fig. 6d). Positive (negative) GDD is often described as "red first, blue later" ("blue first, red later") as shown in Fig. 6b, and is referred to as normal (anomalous) dispersion or a positive (negative) chirp process.

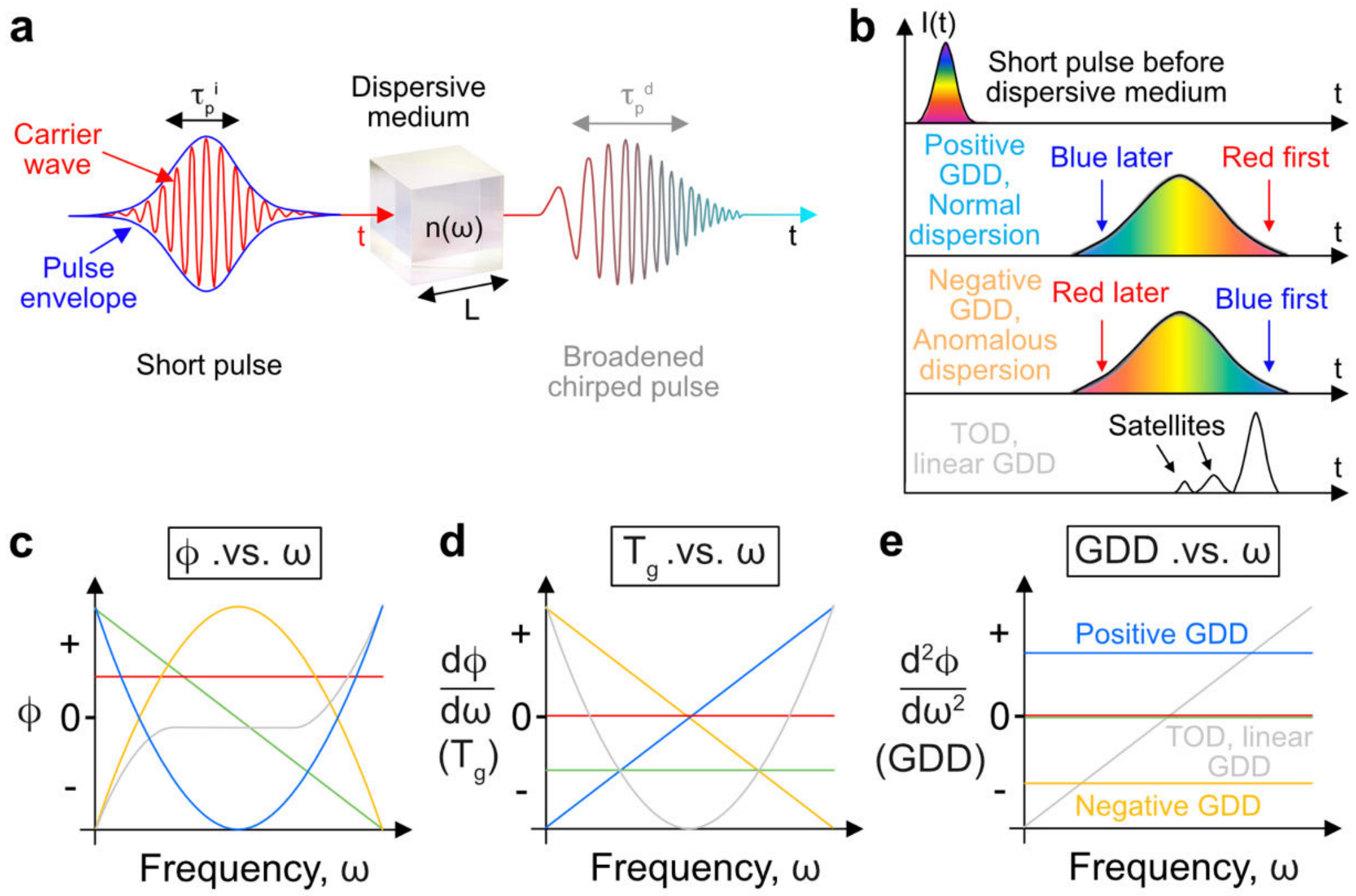

**Figure 6.** Propagation of intense, femtosecond laser pulses. (a) A schematic of a transform limited, short pulse of duration $\tau_{\mathrm{p}}^{\mathrm{i}}$ traversing through a dispersive medium of length $L$ and refractive index $n(\omega)$, generating a broadened, chirped pulse with a longer pulse duration, $\tau_{\mathrm{p}}^{\mathrm{d}}$. The carrier wave (red oscillating distribution) and pulse envelope (blue distribution) of a laser pulse are shown. (b) Frequency content of a laser pulse in time before (top) and after interacting with a dispersive medium (all other panels). The distribution corresponding to positive, negative and linear GDD are shown. (c-e) The phase, $\phi$, (c), group delay, $T_{\mathrm{g}}$, (d), and group dispersion delay, GDD, (e) as a function of frequency, $\omega$, are shown for five pulses with: no chirp or group delay (red), only group delay (green), positive GDD (blue), negative GDD (orange), and TOD (grey).

The fourth term contains the third order phase, TOD $= \frac{\partial^3\phi}{\partial\omega^3}$, which is the second order phase (i.e. $\phi_0$ changes cubically as a function of $\omega$) and is the quadratic variation in the group delay $T_{\mathrm{g}}$ as a function of frequency. The subsequent linear variation in the GDD as a function of $\omega$ (see grey distribution in Fig. 6e) leads to the interference of low and high frequency components, and the appearance of satellites in the dispersed pulse after third order dispersion. Dispersion can be compensated with materials such as BK7, SF10 and sapphire which give positive GDD and TOD values in the near-infrared regime (e.g. sapphire has a GDD and TOD of +58 $fs^2$/mm and +42.19 $fs^3$/mm [164,165]). Negative GDD can be achieved with gratings and prism compressors, with the former capable of larger GDD values. Achieving single-cycle pulses requires matching higher orders (e.g. TOD) than the second order dispersion (i.e. GDD). To avoid the broadening of pulse duration as a result of GDD or TOD, a laser pulse can be (i) propagated in vacuum, (ii) used in combination with highly reflective optics, (iii) propagated through the smallest thickness of material, or (iv) re-compressed using chirped mirrors.

### *2.7 Velocity-map imaging and COLTRIMS reaction microscope spectrometers*

So far, we have seen how we can prepare a molecular ensemble and ionized it using ultrashort laser pulses. An important aspect of CEI and LIED is to detected the three-dimensional momentum distribution of the charged fragments that result from the (multiple) ionization of the molecule. This is typically achieved using either a velocity-map[166] imaging[167] (VMI) spectrometer [166–169] or a cold target recoil ion momentum spectroscopy (COLTRIMS)[170–172] reaction microscope (REMI)[173]. In a VMI spectrometer, the initial particle's velocity distribution is projected into a two-dimensional position- and time-sensitive detector. Fig. 7a shows the VMI concept of charged particle detection. Here, the laser pulse (LP; pink) intersects the molecular beam (MB; grey) to generate charged particles (red circles) with a three-dimensional velocity distribution (black arrows) between two charged repeller and extractor electrode plates (blue). Tuning the voltage ratios between the two electrodes provides a scenario where charged particles with the same velocity can be projected onto the same position and at the same time on the detector irrespective of their starting position in the interaction region. Once the charged particle impinges onto the front microchannel plate (MCP) of the detector assembly, a flash of light is emitted from the back of a scintillator (e.g. phosphor) screen which is subsequently measured with a camera detector. Other types of detector assemblies can also be used, some of which will be discussed in the next sub-section, each possessing their own advantages and challenges. There are many different variations of the VMI concept in electron-ion spectrometers which employ, for example, the use of additional electrodes [174], an einzel lens [175], the pulsing of an electrode [176], and the injection of a target gas jet into the repeller plate through a capillary [177]. In a VMI spectrometer, a two-dimensional (2D) projection of the 3D velocity distribution is measured. If the system under study possesses cylindrical symmetry then the initial 3D momentum distribution can be reconstructed by an Abel inversion procedure such as the pBasex [178]. In general, the open electrode designs of VMI spectrometers achieve detection of charged particles in the entire solid angle of $4\pi$. However, typical VMI set-ups employ quite high extraction fields to achieve $4\pi$ detection and for the charged particle to acquire enough kinetic energy to be detected by the microchannel plate with enough efficiency.

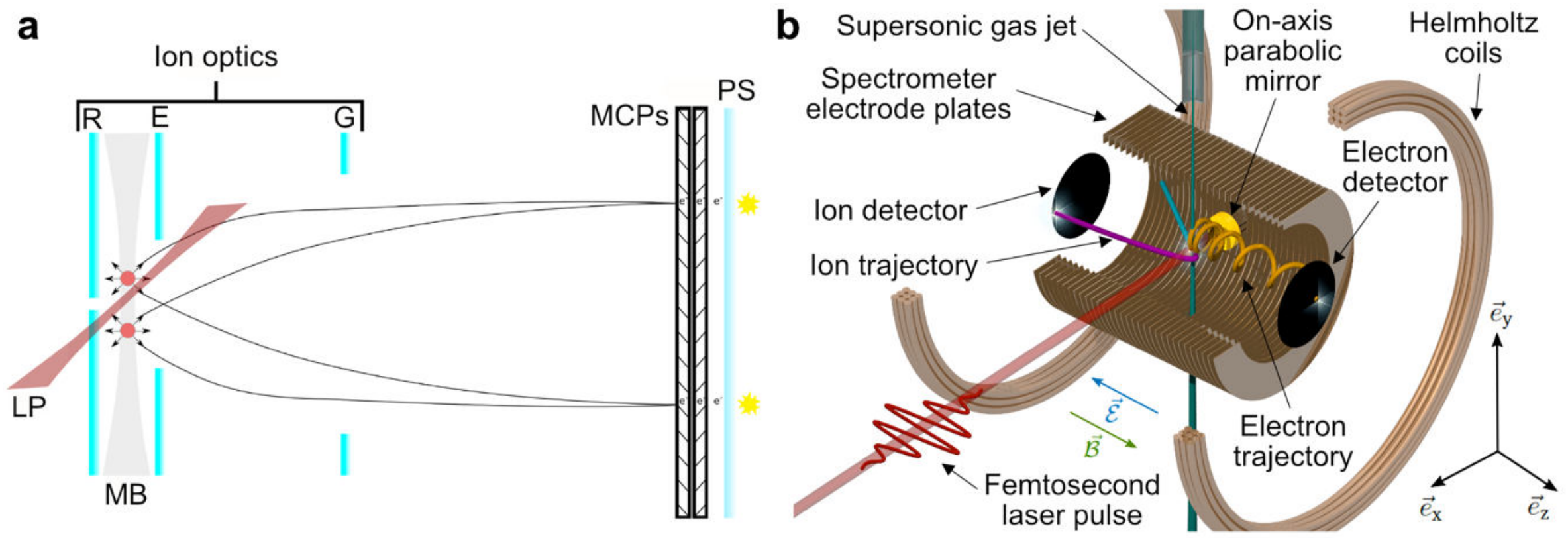

**Figure 7.** Concept of velocity-map ion-imaging and COLTRIMS reaction microscope spectrometers. (a) A molecular beam (MB; grey shaded) jet of gaseous molecules (red circles) is ionized by a laser pulse (LP; red shaded). The generated ion fragments are extracted from the interaction region between the charged repeller (R) and extractor (E) electrode plates through a grounded (G) electrode towards a position- and time-sensitive detector consisting of microchannel plates (MCPs) and a phosphor screen (PS). A flash of light is generated by the PS and detected by a camera. (b) Schematic of a COLTRIMS reaction microscope spectrometer. A supersonic gas jet (green) is overlapped with a femtosecond laser pulse (red) that is focussed into the interaction region with an in-vacuum on-axis gold-coated paraboloid. The generated charged ions and electrons are extracted towards their corresponding detectors with electrostatic and magnetic fields, respectively, using spectrometer electrode plates and Helmholtz coils. The detectors are composed of microchannel plates (MCPs) and delay line anodes (DLAs). Panels (a) and (b) are adapted from Ref. [69] and Ref. [154], respectively.

Advanced ion-electron momentum spectrometers such as the cold target recoil ion momentum spectroscopy (COLTRIMS)[170–172] and reaction microscope (REMI)[173] spectrometer employ much lower extraction fields and can directly measure the full three-dimensional momentum distribution of charged particles with sub-10 meV momentum resolution under specific instrument parameter settings. It should be also noted that the resolution of a COLTRIMS and REMI set-up is not constant over the full kinetic energy range, and has a typical energy resolution $\Delta E/E$ of <2%. An additional important advantage of the COLTRIMS and REMI set-up is the capability to perform kinematically complete coincidence measurements of the 3D momenta of the charged particles (e.g. ions and electrons) that are generated at the focus of the intense laser field. Under these coincidence conditions, charged particles are detected at a count rate of <<1 event/shot to ensure that the generated particles originate from the fragmentation of a single molecule. Specifically, this is achieved by examining the detected momentum distributions of charged particles to ensure that their momentum sum is conserved. This allows the direct measurement of the full three-dimensional momentum distribution of particles without the requirement of cylindrical symmetry nor the use of inversion algorithms as is typically required in VMI. Fig. 7b shows a typical COLTRIMS REMI set-up [155,179]. Here, a laser pulse is focussed onto a supersonic gas jet using an on-axis gold-coated parabolic mirror. The molecular beam is collimated using two skimmers prior to entering the interaction region. The ions (pink, Fig. 7b) and electrons (yellow, Fig. 7b) are projected towards their corresponding separate detector assemblies consisting of an MCP and delay-line anode set-up (see next sub-section for more information on delay-line anode detectors). In the case of the positive ions, the spectrometer plates are negatively charged, with the homogenous electric field only required to extract the ion towards the detector. In the case of electrons, a combination of a magnetic field (generated by the Helmholtz coils) and a positively charged homogeneous electric field (generated by spectrometer plates) forces the electrons to travel with a cyclotronic trajectory towards the electron detector. The radius of the cyclotron trajectory is dependent on the transverse momentum of the electron. In the absence of an external magnetic field, electrons with a variety of transverse momenta are detected at the same position on the detector. This has important implications for LIED since the re-scattering process in the mid-infrared regime can generate electrons possessing different transverse momenta which encodes molecular structure information.

### *2.8 Fast imaging detection of 3D velocity distributions*

The 3D velocity distribution of charged particles can be detected using fast imaging position- and time-sensitive detectors which can provide the position $(x, y)$ and the time of flight, $t$, of a charged particle. The simultaneous measurement of multiple fragments of different mass-to-charge ratio, $m/q$, under the same experimental conditions using $t = k\sqrt{m/q}$ where $k$ is a constant related to the instrument settings. Two types of detectors are commonly used: low-count rate delay-line anode (DLA) detectors [180–183] and high-count rate CMOS-based sensor detectors [184–194]. The DLA detector consists of two or more wound wires biased with a constant voltage which are located behind microchannel plates (MCPs). When an electron cloud emitted from the MCPs impinges on the DLAs, the voltage of each wire is reduced. By monitoring the timing differences and sums induced by the changes in voltage in each delay line the position and arrival time of the particle can be retrieved with a precision of 100 μm and 100 ps. DLA detectors are limited to only a few hits per laser shot, and are well-suited for low-count rate (i.e. <<10 events/shot) measurements performed at a high repetition rate (i.e. >1 kHz) such as those in coincidence spectroscopy. CMOS-based detectors are better suited for relatively high-count rate (i.e. >> 10 events/shot) measurements (i.e. <1-kHz). CMOS-based sensors consist of a large number of "intelligent" pixels that can be read out separately to provide the position of the incoming particle on the detector. The timing information or arrival time of the particle with respect to an external trigger is retrieved typically using an on-board counter in each pixel that increases whenever a particle impacts the pixel in the sensor. The timing information can be stored in the on-board electronics of each pixel until the next laser shot, such as in the PImMS [184–191] or TimePix [192–194] sensors, or it can be read dynamically during the same acquisition window, such as in the TimePix3 [195–197] sensor. These sensors can be implemented into an out-of-vacuum lens-based camera system, for example, with PImMS [186,189] and TimePix [198–201] cameras. The latest generation of PImMS and TimePix sensors (PImMS2 and TimePix3) have a timing resolution of 12.5 ns and 1.5 ns, respectively. The nanosecond barrier is expected to be surpassed by the TimePix4 [202] sensor which will provide the ability to better resolve the detected time-of-flight distribution of charged particles which typically span tens or hundreds of nanoseconds for electrons and ions, respectively.

## 3. Coulomb explosion imaging

In femtosecond Coulomb explosion imaging (CEI) [61–102], multiple electrons are ejected from the molecule on a timescale short enough (< 1 fs [62–64,203]) where molecular motion (e.g. vibration and rotation) is frozen during the CEI process to generate a highly charged molecular ion [66]. The resulting intramolecular Coulomb repulsion between the charged sites far outweigh the binding energy of the unstable molecular ion. This results in the rapid break-up of the molecular ion which generates multiple ionic fragments. The trajectory of the ionic fragments can be modelled using classical Newtonian laws of motion within a Coulomb potential since the forces governing the Coulomb explosion process are purely Coulombic in nature. The positions of atoms and functional groups in the molecule at the time of Coulomb explosion is encoded in the three-dimensional velocity distribution of the charged fragments measured using a VMI or COLTRIMS and REMI spectrometers. Correct structural identification requires that the molecular structure does not change significantly during the Coulomb explosion process, abiding by the axial recoil approximation [74]. In this approximation, the Coulomb exploding fragment ions recoil along the original bond axes of their parent molecule. Within this approximation, the molecular structure can be measured by correlating the emission directions of the ionic fragments from their measured momenta. Further analysis procedures have enabled CEI measurements under conditions where the axial recoil approximation is not met [74]. Moreover, combining CEI with coincidence [170,171,173,204–206] or covariance detection [68–73,75,78,80,83,84,101,204,207–211] leads to improved accuracy of the retrieved molecular structure, which will be discussed in further detail later in this section.

In the first CEI studies, a fast molecular ion beam passed through a thin metal "stripping" foil that subsequently removed one or more valence electrons from the singly-charged molecular ion and generated a multiply charged molecule that Coulomb exploded on the sub-femtosecond timescale [61–64,203], with the resulting fragments subsequently detected. However, the arrival time of the molecular beam at the foil can only be synchronized on picosecond or longer timescales, which is too slow for femtosecond-resolved measurements of molecular dynamics.

The Coulomb explosion of molecules can be achieved using laser pulses of a wide-range of frequencies. In the optical spectral range, high intensities ($> 10^{14}$ W/cm$^2$) are required to multiply ionized a molecule and observe the Coulomb explosion of the molecular ion, which can be achieved with current femtosecond laser technology. In contrast, short wavelength radiation sources provide a means to multiply ionize a molecule at already modest intensities (> $10^{12}$ W/cm$^2$) since the removal of an electron from a core orbital is generally, if not always, accompanied with an Auger process, which leads to the emission of a second electron. As such, a multiply charged molecular ion is formed that can Coulomb explode with the caveat that the Auger process should be fast enough such that the structure of the molecule remains unchanged during the Coulomb explosion process. Ultrashort pulses of short wavelength radiation that can be used not only for structural determination but for dynamical studies are only available since the last decades at free-electron laser (FEL) facilities, such as the European X-ray free-electron laser (Eu-XFEL) and LCLS [149,212,213], as well as through the high harmonic generation process [105,109,121,160–162].

### 3.1 Kinetic energy release mapping of molecular structure and dynamics

The measured kinetic energy release (KER) of the ionic fragments, with corresponding centre-of-mass located at position $i$ and $j$ at the time of ionization, is directly related to the scalar distance, $R_{\mathrm{ij}}$, between two ionic fragments by[66]

$$E = k_e \frac{q_1 q_2 e^2}{2R_{\mathrm{ij}}}, \qquad (12)$$

where $q_1$ and $q_2$ are the charges of the ion fragments, $k_\mathrm{e}$ is the Coulomb's constant, and $e$ is the electron charge. Using Eqn. 12, the internuclear distance between the different atoms in a molecule can be retrieved from a measurement of the kinetic energy of the charged particles that have repelled by the Coulomb force. This is shown in Fig. 8 by the relationship between KER and I-I internuclear distance, $R_{\mathrm{II}}$, in the time-resolved pump-probe Coulomb explosion of dissociating $I_2$. Laser-induced Coulomb explosion imaging was first demonstrated by Stapelfeldt *et al.* [66] to monitor the wave packet dynamics of dissociative $I_2$ molecule. Using a first femtosecond laser pulse, a wavepacket was launched in the $\mathrm{A}^3\Pi_{1\mathrm{u}}$ dissociative excited state of the molecule. The dissociation dynamics was directly mapped in the $I^+$ KER resulting from strong field ionization of the dissociating $I_2$ molecule followed by Coulomb explosion, as shown in Fig. 8. At increasing pump-probe delays, and thus increasing $R_{\mathrm{II}}$ internuclear distance between $I^+$ and I, a lower $I^+$ KER is measured. For example, the $I^+$ KER decreases from 1.2 eV at 0.4 ps to 0.7 eV at 1.1 ps, which is a result of the increase in $R_{\mathrm{II}}$ from ~7 Å to ~14 Å, respectively, as shown in Fig. 8b.

The first experiments employing CEI were limited to simple diatomic molecules, in which the measurement of the KER of a single charged fragment is sufficient to retrieve the change in the internuclear distance of the diatomic molecule.

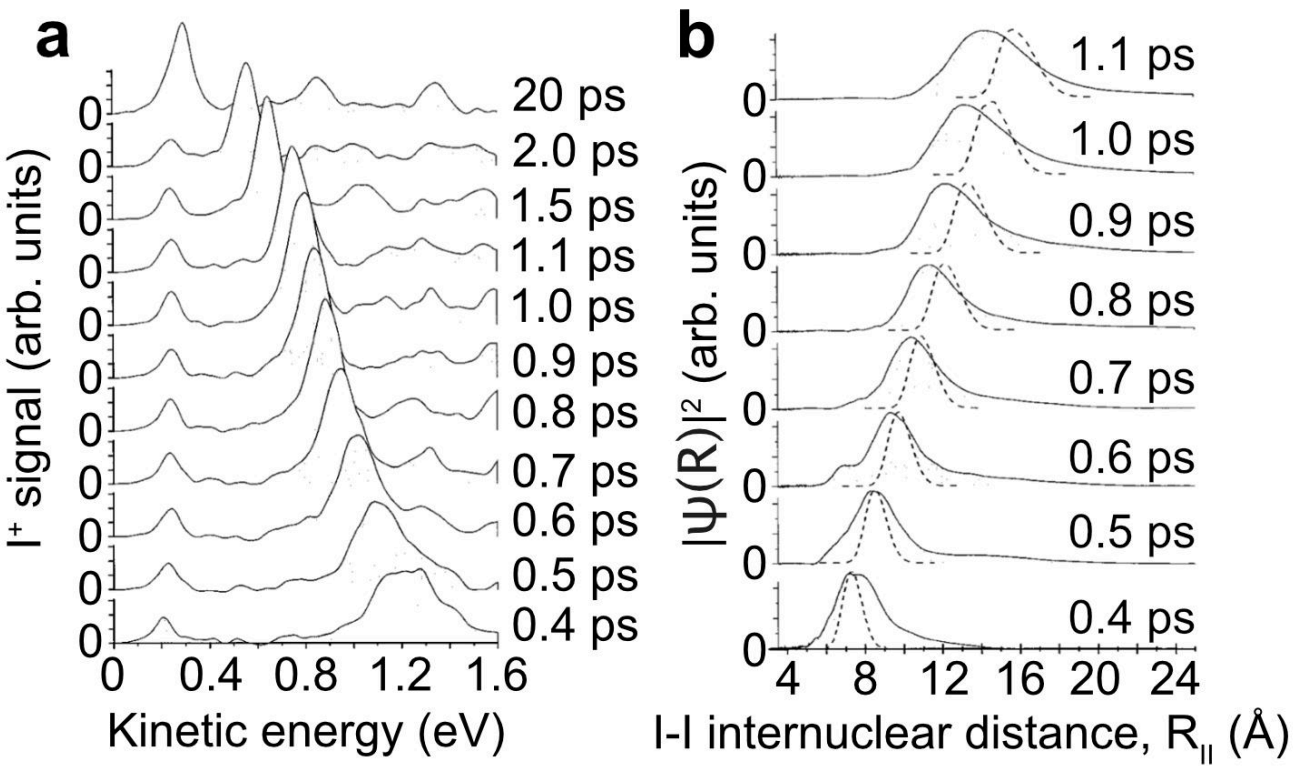

**Figure 8.** Dissociative ionization dynamics in $I_2$ probed by CEI. (a) Kinetic energy release (KER) spectrum of $I^+$ and (b) the square of the nuclear wave function, $|\psi(R_{\mathrm{II}})|^2$, of dissociating $I_2$ as a function of I-I internuclear distance, $R_{\mathrm{II}}$, measured at various pump-probe delays. Figure adapted from [66].

For larger polyatomic molecules, the simultaneous detection of fragments with different mass is generally required to image the molecular structure and dynamics, which can be achieved using more advanced detectors with multi-hit capabilities such as the PImMS sensor (see Fig. 9). The mass-to-charge spectrum (Fig. 9a) shows the various ionic fragments generated following the one-photon ultraviolet (UV) photodissociation of the C-I bond in $CH_2IBr$ and its subsequent Coulomb explosion by an intense 800-nm pulse. In addition, the $I^+$ and $CH_2Br^+$ momentum distribution measured at two pump-probe delays are shown in Fig. 9b. Here, there exist different VMI rings that can be assigned to various dissociative ionization channels (at low momenta) and Coulomb explosion channels (at high momenta). Increasing the pump-probe delay leads to the reduction in kinetic energy of the high-energy ion fragments as shown by the white arrow in Fig. 9b. A two-dimensional map of the $CH_2Br^+$ ion fragment's kinetic energy as a function of pump-probe delay is shown in Fig. 9c. The KER of the high-energy Coulomb exploding $CH_2Br^+$ fragment decreases from 2 eV to 0.6 eV with increasing delay (0 ps to 2.65 ps) due to the C-I bond length elongation following UV dissociation (e.g. see feature II in Fig. 9c). Similar effects are observed in the UV dissociation of $CH_2ICl$ and $CH_3I$ (see Figs. 9d and 9e) but with the minor difference that the high-energy dissociating channel is split into two contributions. This effect can be well explained if we consider the spin-orbit splitting of the iodine atom following the dissociation of the C-I, yielding either the excited $I^*(^2P_{1/2})$ or ground $I(^2P_{3/2})$ spin-orbit states prior to Coulomb explosion. The KER of the two spin-orbit channels were confirmed by a simple classical Newtonian trajectory propagation calculation which modelled the photodissociation and Coulomb explosion process. Specifically, the internuclear distance between the two fragments (e.g. $CH_2Cl$ + I), and thus the KER of each fragment using Eqn. 12, were calculated as a function of pump-probe delay. During the photodissociation and Coulomb explosion steps, the ($q_1$,$q_2$) charges of (0,0) and (+1,+1) were assumed, respectively, where $q_1$ and $q_2$ are the charges of $CH_2Cl$ and I. In general, a good agreement between measured and calculated data are observed in Figs. 10c-10e. In all three molecules, there also exists a high energy channel at all delays which corresponds to the undissociated molecule that is Coulomb exploded to generate $CH_2X^+$ + $I^+$ (e.g. see feature I in Fig. 9c) where X is a halogen or hydrogen atom. A low-energy channel also exists in all three spectra (e.g. see feature III in Fig. 9c), arising from the dissociation of the monocation parent molecule which generates the $CH_2X^+$ ion and a neutral I atom after interaction with the pump and probe pulses. In general, as demonstrated above, tracking changes in the KER of an ionic fragment reveals information about timescales and key bond lengths involved in a photochemical reaction.

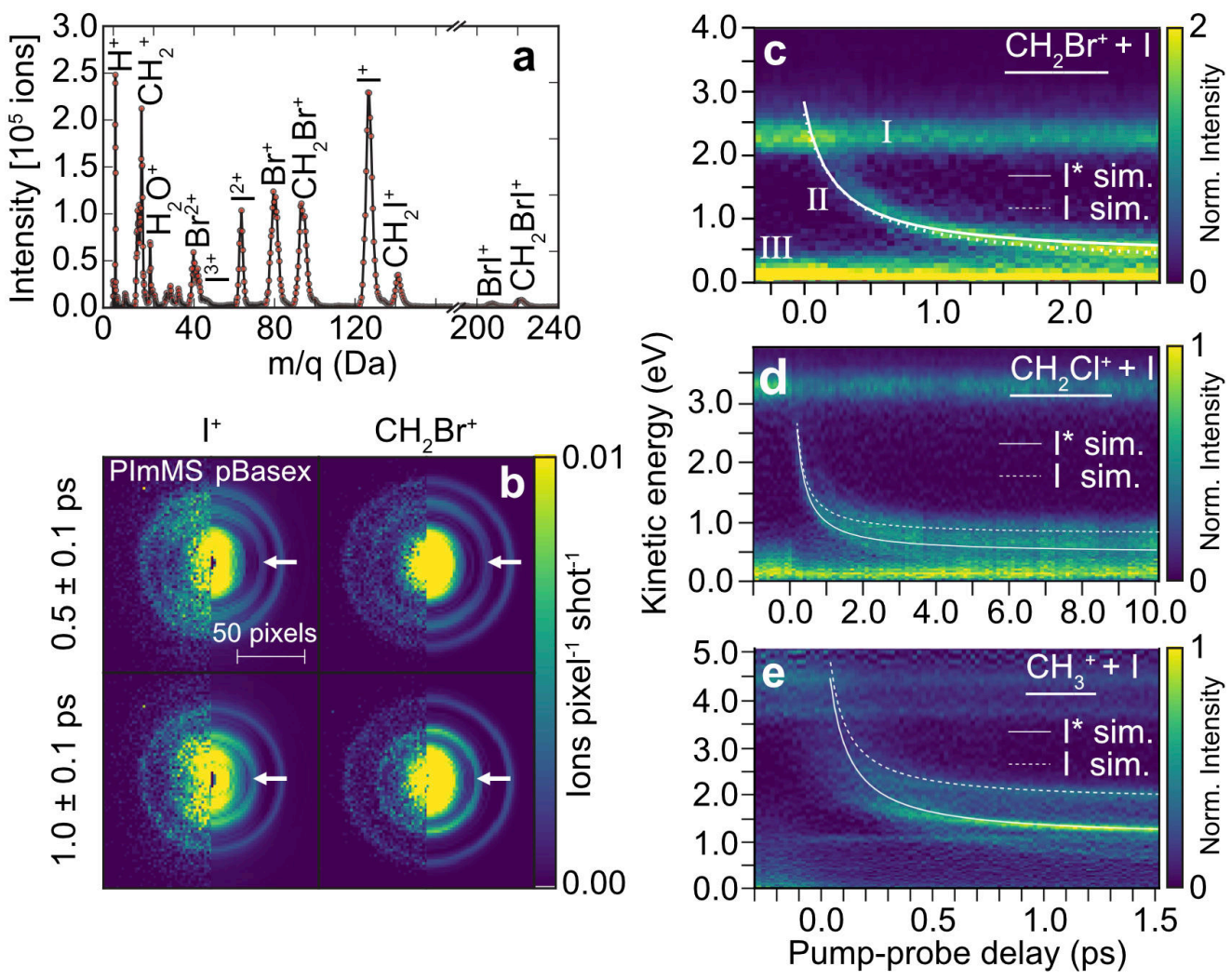

**Figure 9.** Multi-mass CEI studies of photodissociation dynamics in halogenated methanes. (a) Mass-to-charge spectrum of ion fragments generated following the UV photodissociation of the C-I and C-Br bonds in $CH_2IBr$ at 272 nm subsequently probed by CEI at 800 nm. (b) The VMI images of $I^+$ (left) and $CH_2Br^+$ (right) at 0.5 ps (top) and 1.0 ps (bottom) pump-probe delays. The white arrow indicates the reduction of the Coulomb explosion channel in kinetic energy with increasing delay. The raw PImMS and pBasex-inverted data are shown. (c)-(e) The two-dimensional kinetic energy distributions as a function of pump-probe delay for $CH_2Br^+$ (c), $CH_2Cl^+$ (d) and $CH_3^+$ (e) ion fragments following the UV-pump CEI-probe of $CH_2IBr$ (c), $CH_2ICl$ (d), and $CH_3I$ (e). The simulated data for I* and I are shown as solid and dashed lines. Panels (a)-(c) adapted from [90], and panels (d)-(e) adapted from [80].

### *3.2 Covariance and coincidence mapping of CEI data*

It is desirable to image the complete three-dimensional (3D) molecular structure in real-space beyond only measuring the one-dimensional scalar distances between two ion fragments (as shown in Fig. 8 and Eqn. 12). Doing so enables the complete 3D molecular structure to be determined. This can be achieved by combining CEI with advanced analytical methods such as coincidence [170,171,173,204–206] and covariance [68–73,75,78,80,83,84,101,204,207–211] analysis. In both methods, the emission direction of two (or more) ion fragments are correlated to each other to reveal their relative positions within the molecule. This is achieved by plotting one (or more) of the ion fragments in the recoil-frame of a second reference ion, as illustrated in Figs. 10a – 10c. In coincidence analysis, the coincidence between two particles $X$ and $Y$ are given by [204]

$$Coinc(X,Y) = \sum_{i=1}^{i=N} X_i Y_i \,, \tag{13}$$

where $i$ is the laser shot number and $N$ is the total number of laser shots. The count rate must be kept significantly below one event per acquisition shot to avoid the appearance of falsely correlated events, called "false coincidences", which can appear at relatively high count rates (i.e. >> 0.01 per shot). This ensures that the measured ion fragments originate from a single molecule and are correlated. Fig. 10d shows the appearance of "false coincidences" at high count rates (i.e. >> 0.01 per shot), which are due to the detection of two uncorrelated ions, and can be corrected for by covariance analysis.

The concept of covariance imaging is based on utilizing the correlated fluctuations typically occurring in an experiment. For example, fluctuations in the time-of-flight intensity of two particles can be correlated with fluctuations in the laser intensity. Since gas-phase measurements follow Poisson statistics, the uncorrelated false coincidences, $\langle X\rangle\langle Y\rangle$, can be subtracted from the correlated signal, $\langle XY\rangle$, in a shot-to-shot and event-based analysis procedure to generate a covariance map of two ions $X$ and $Y$, [70,204,207,209]

$$Cov(X,Y) = \langle XY\rangle - \langle X\rangle\langle Y\rangle \,, \tag{14}$$

where $\langle X\rangle$ is the average number of counts for particle $X$ as given by [70,204,207,209]

$$\langle X\rangle = \frac{1}{N}\sum_{i=1}^{i=N} X_i \,. \tag{15}$$

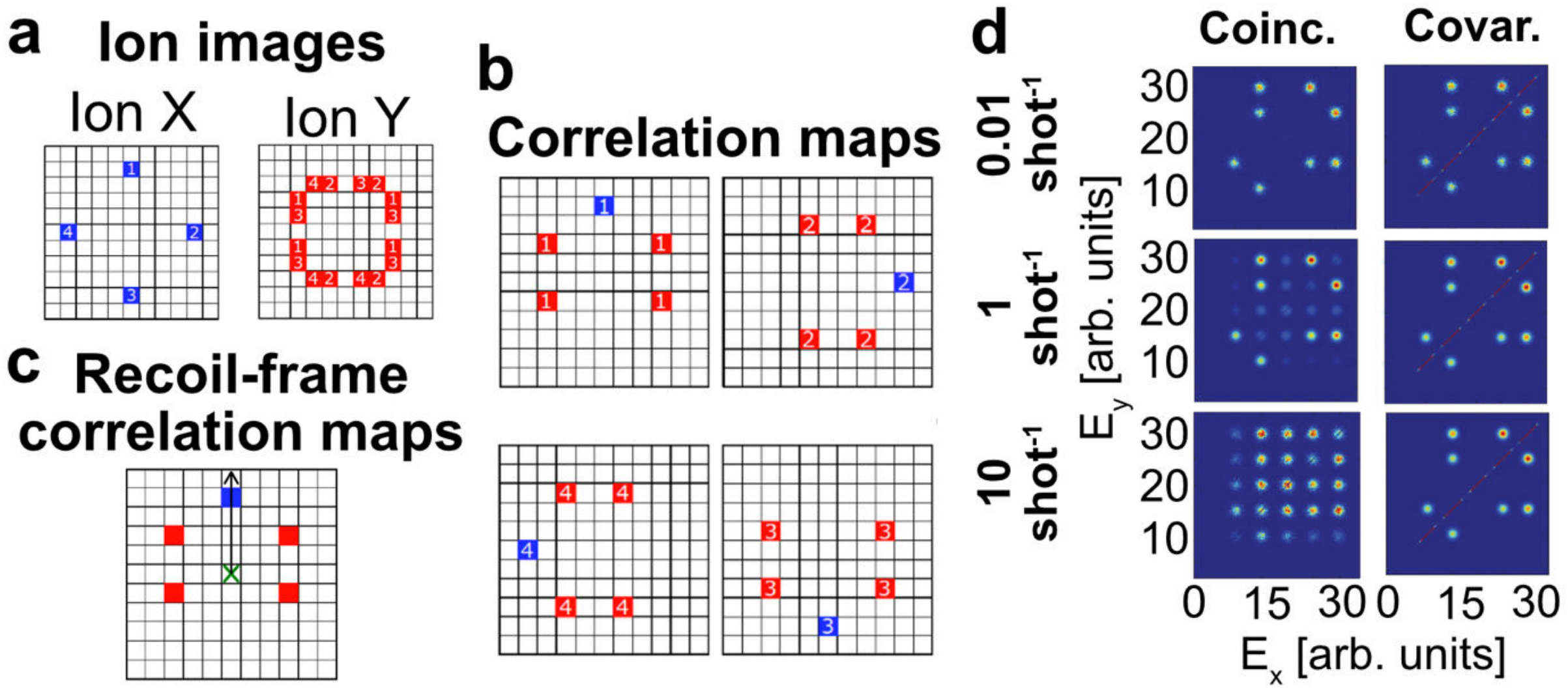

**Figure 10.** Coincidence and covariance imaging. (a) Two ion fragments are identified by their corresponding time of flight ranges. Particles $X$ and $Y$ are termed the "reference" and "plotted" ions, and their laboratory-frame VMI images are shown summed over four acquisition shots. (b) Correlated maps for each individual acquisition shot. In coincidence analysis, this simply corresponds to the VMI image for each shot being shown. In covariance analysis, the covariance between each pixel of the reference ion VMI image to each pixel of the plotted ion VMI image is calculated and shown for each shot. (c) The correlation map in each shot is rotated into a common recoil-frame such that the velocity vector of the reference ion $X$ (blue square) is plotted along the black arrow, with the plotted ion $Y$ (red squares) shown relative to the reference ion $X$. (d) Two-fold correlation maps at three count rate conditions of 0.01, 1 and 10 counts per shot retrieved by coincidence and covariance analysis methods. Panels (a-c) and (d) adapted from [69] and [204], respectively.

Fig. 10d shows the ability of the covariance method to correct for the detection of falsely correlated signal as demonstrated by the absence of false coincidences in the covariance maps in the high-count rate regime (>> 1 event per shot). In the very low count rate regime (<< 1 event per shot), the uncorrelated false coincidence term $\langle X\rangle\langle Y\rangle$ becomes negligible and so $Cov(X,Y)$ and $Coinc(X,Y)$ are directly comparable through[204,209]

$$Coinc(X,Y) = NCov(X,Y). \quad (16)$$

Since coincidence imaging is performed under low-count rate conditions, coincidence measurements are only worthwhile if they are performed at sufficiently high repetition rates (i.e. >> 10-Hz) to measure statistically significant data within a reasonable integration time. The low count rate multi-hit capability of DLA detectors make it well-suited for coincidence measurements, although the Tpx3Cam CMOS-based camera has also been employed in low count rate coincidence VMI measurements at 1-kHz [201]. Covariance imaging is most beneficial in the high-count rate regime, and as such CMOS-based PImMS and Timepix sensors are well-suited given their capability to measure $(x, y, t)$ data for many hundreds of events per acquisition shot. Measurements at such high-count rates is usually performed at relatively low repetition rates (<1 kHz) to avoid data buffering and throughput issues due to the vast volume of data generated in such sensors. In general, covariance measurements can acquire correlated data with an integration time two-orders-of-magnitude lower than in coincidence measurements as demonstrated in the strong-field ionization measurements of $D_2O$ at the same repetition rate [102]. However, it should be noted that a fundamental requirement exists for performing covariance measurements: a stable optical or FEL laser system possessing correlated (and not random) fluctuations in laser parameters that generates correlated changes in the measured ion signal. The self-amplified spontaneous emission (SASE) nature of FELs leads to new, additional background signal that contributes to the full covariance signal. Partial covariance has been demonstrated as a tool to retrieve correlated signal from SASE-generated ion ToF signals. In partial covariance mapping, [209,214,215] an additional contribution is subtracted from the full covariance calculation which is related to the fluctuations arising from the SASE process of FELs. Incorporating FEL beamline data (e.g. pulse energy, pulse pointing, pulse duration etc) into current covariance analysis algorithms will lead to advanced partial covariance algorithms suitable for FEL-based VMI measurements, enabling the extraction of recoil-frame covariance data generated by FELs. It should also be noted that laser-induced alignment of molecules[76,216,217] will further improve the capability to extract correlated signals from VMI measurements, but it is not strictly required.[69]

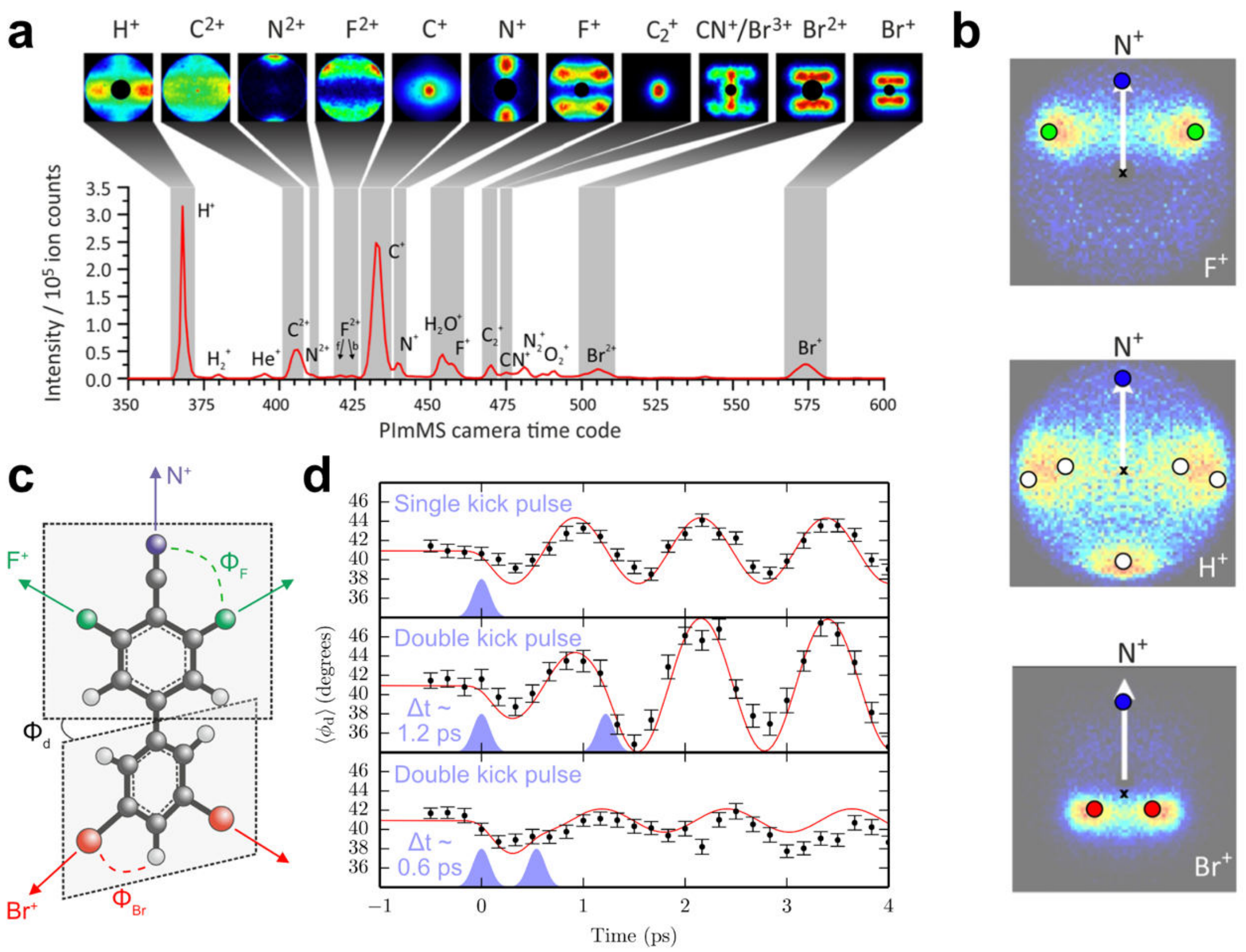

**Figure 11.** Structural identification and vibrational dynamics probed by CEI and covariance mapping. (a) Time-of-flight spectrum and corresponding VMI images following the Coulomb explosion of BFCbP molecules with an intense 800-nm femtosecond pulse. (b) Recoil-frame covariance maps of F+, H+ and Br+ plotted with the N+ reference ion. Simulated covariance-mapped CEI data are overlaid on top of the measured data as circles. (c) Molecular structure of BFCbP. The dihedral angle, $\phi_d$, and the angle of the fluorine and bromine atom with respect to the $C_2$ rotational axis along the C-N bond, $\phi_F$ and $\phi_{Br}$, respectively, are shown. (d) Time-resolved response of $\phi_d$ after the excitation of BFCbP molecule by a single (top) or double kick pulses. In the latter, double kick pulses were sent at two delays between the first kick pulse and CEI probe pulse of 1.2 ps (middle) and 0.6 ps (bottom) corresponding to a full and half vibrational period of BFCbP, respectively. Panel (a) was adapted from [71], panels (b) and (c) are adapted from [84] and [72], and panel (d) is adapted from [72].

### *3.3 3D CEI with fast imaging sensors*

Typical examples of CEI combined with covariance mapping is presented in this section. In the first experiment, 3,5-dibromo-3′,5′-difluoro-4′-cyanobiphenyl (BFCbP) molecules were exposed to an intense femtosecond 800-nm laser pulse. The ionic fragments were recorded using a PImMS fast imaging sensor composed of 72 × 72 pixels and a timing resolution of 12.5-ns. Fig. 11A shows the time-of-flight (ToF) spectra and a selection of momentum distributions of specific fragments.[71] From the 2D slice through the 3D momentum distribution of $N^+$, $F^+$, $H^+$ and $Br^+$ ions, a shot-to-shot covariance analysis yields the corresponding $Cov(N^+, F^+)$, $Cov(N^+, H^+)$ and $Cov(N^+, Br^+)$ recoil-frame covariance maps in the recoil-frame of the $N^+$ reference ion (see Fig. 11b).[70,84] The result of trajectory simulations using a simple classical Newtonian trajectory propagation model is superimposed on top of the measured covariance maps which show a very good agreement. For example, the simulated data reproduce the forward and backward emission direction of $F^+$ (green circles, top panel Fig. 11b) and $Br^+$ (red circles, top panel Fig. 11b) ions, respectively, relative to the $N^+$ ion (blue circle) shown in the calculated $Cov(N^+, F^+)$ and $Cov(N^+, Br^+)$ covariance maps agree well with the measured data. Moreover, the position of the hydrogen atoms (white circles, middle panel Fig. 11b) is also well reproduced in the $Cov(N^+, H^+)$ map.

In a second measurement, an additional femtosecond "kick" laser pulse was used to excite the torsional vibrational motion between the two phenyl rings in BFCbP, leading to a change in the dihedral angle, $\phi_d$, between the two phenyl rings. The ensuing vibrational motion dynamics was then probed by a femtosecond CEI probe laser pulse.[72] For this experiment, an additional nanosecond laser pulse was used to align the molecule along its $C_2$ rotation axis parallel to the C-N bond. Laser alignment of molecules provides the advantage to limit the rotation of gas-phase molecules by fixing one or more molecular axis, which maximizes the signal-to-noise quality of the measured CEI data. The delay between the excitation and probe laser pulses is varied and the fragments arrival time and KER were recorded with the PImMS detector. Since the dihedral angle, $\phi_{\rm d}$, is equivalent to the relative angle $\phi_{\rm F} - \phi_{\rm Br}$ between the emission direction of the $Br^{+}$ and $F^{+}$ fragments, this gives $\phi_{\rm d} = \phi_{\rm F} - \phi_{\rm Br}$. The dihedral angle can therefore be directly extracted from the $Cov(\mathrm{Br}^{+}, \mathrm{F}^{+})$ covariance map between the $Br^{+}$ and $F^{+}$ fragments which images changes in the relative angle between the two fragments, $\phi_{\rm F} - \phi_{\rm Br}$. The time-resolved changes in the dihedral angle with a single or double kick pulse is shown in Fig. 11d. In the case of the single kick pulse case (top panel, Fig. 11d), an oscillatory behaviour in the dihedral angle with a period of ~1.2 ps and an amplitude of 3° is observed. The oscillatory behaviour is further enhanced or reduced by using a second kick pulse that arrives either a full (~1.2 ps) or half (~0.6 ps) a vibrational period, respectively, after the first kick pulse (middle and bottom panels, Fig. 11d, respectively). Therefore, using two precisely timed kick pulses, the vibrational motion of the molecule can be amplified or quenched.

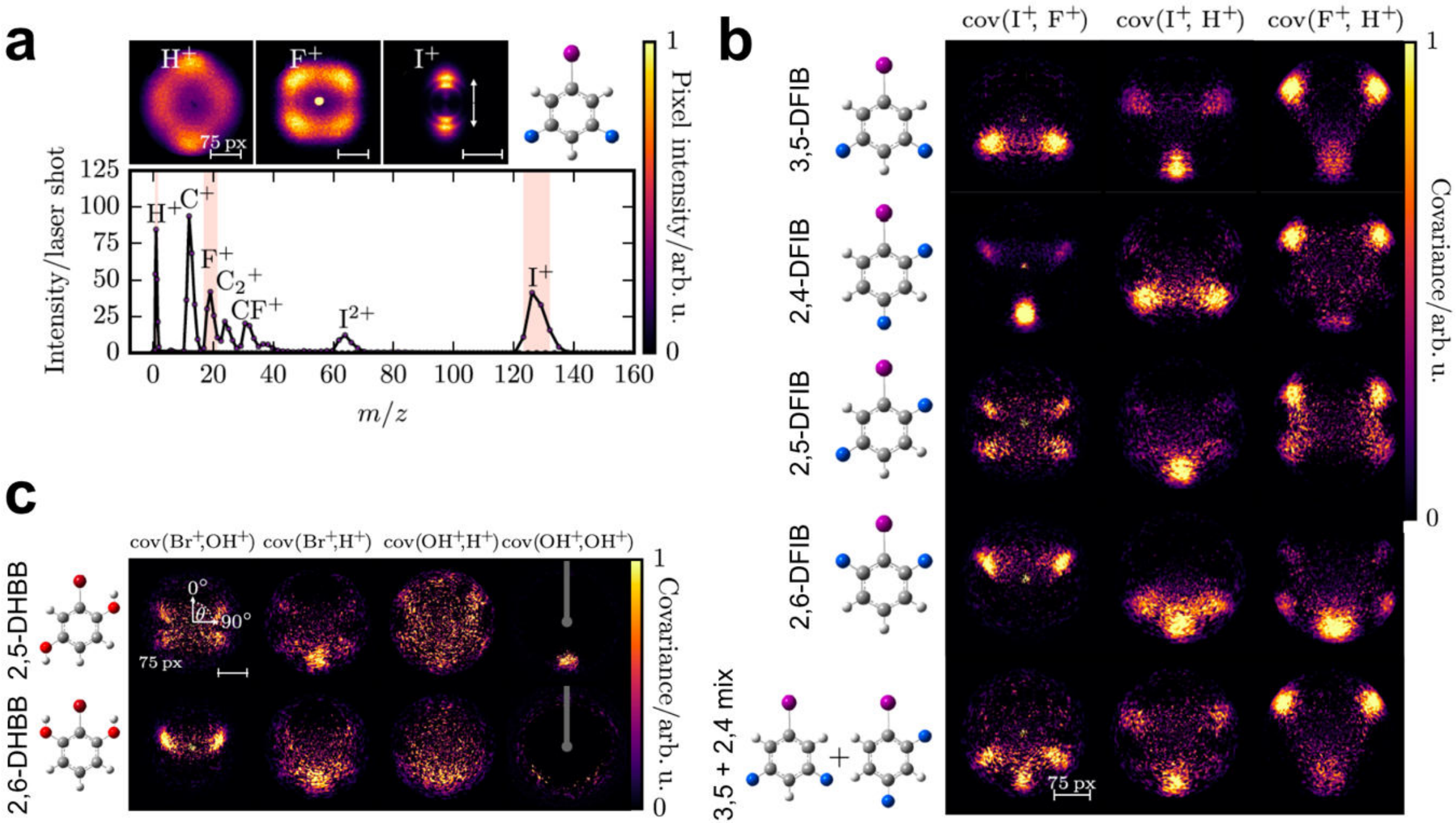

**Figure 12.** Isomer identification of DFIB and DHBB molecules by covariance mapping. (a) ToF spectra and selected VMI images of Coulomb exploded 3,5-DFIB molecules by an intense 800-nm femtosecond pulse. (b) Recoil-frame covariance maps of 3,5-, 2,4-, 2,5-, 2,6-DFIB pure samples, and a mixed sample of 3,5- and 2,4-DFIB. (c) Recoil-frame maps of 2,5-DHBB and 2,6-DHBB. Figure adapted from [69,78].

Covariance imaging coupled with CEI can also be used between different isomers, as shown in Fig. 12 for difluoroiodobenzene (DFIB) and dihydroxybromobenzene (DHBB).[69,78] Fig. 12a shows the ToF spectrum and VMI images of Coulomb exploded $H^{+}$, $F^{+}$ and $I^{+}$ ions following the Coulomb explosion of the 3,5-DFIB molecule measured with the $324 \times 324$ pixel PImMS2 sensor. The $Cov(\mathrm{I}^{+}, \mathrm{F}^{+})$, $Cov(\mathrm{I}^{+}, \mathrm{H}^{+})$, and $Cov(\mathrm{F}^{+}, \mathrm{H}^{+})$ covariance maps for the 3,5-, 2,4-, 2,5-, and 2,6-DFIB isomers are shown in Fig. 12b. From these covariance maps, the identity of the isomer can be unambiguously identified. For example, in the $Cov(\mathrm{I}^{+}, \mathrm{F}^{+})$ and $Cov(\mathrm{I}^{+}, \mathrm{H}^{+})$ covariance maps of the 3,5-DFIB isomer, the $\mathrm{F}^{+}$ and $\mathrm{H}^{+}$ ions are emitted at $(135°, 215°)$ and $(45°, 180°, 315°)$, respectively, to the $\mathrm{I}^{+}$ reference ion, reproducing the 3,5-DFIB molecular structure with good accuracy. The $Cov(\mathrm{I}^{+}, \mathrm{F}^{+})$ map of the 2,4 and 2,5 isomers have additional contributions of the $\mathrm{F}^{+}$ ion at $(315°)$ and $(135°, 315°)$ since the DFIB molecule is free to rotate around its $C_2$ rotational axis along the C-I bond. An additional measurement was performed on a mixture of two isomers consisting of 3,5-DFIB and 2,4-DFIB in a 2.5:1 ratio. Combining the covariance maps of the pure 3,5- and 2,4-isomer samples give a unique $Cov(\mathrm{I}^{+}, \mathrm{F}^{+})$ map that can only have arisen from the mixture of these two isomers, which is in good agreement with the measured mixture data. Fig. 12c shows the covariance maps for two isomers of DHBB, which can be clearly identified from one another with clear signatures particularly seen in the $Cov(\mathrm{Br}^{+}, \mathrm{OH}^{+})$ map that unambiguously identify the 2,5- and 2,6-DHBB isomers in a similar manner to the corresponding $Cov(\mathrm{I}^{+}, \mathrm{F}^{+})$ of the 2,5- and 2,6-DFIB isomers. This demonstrates the sensitivity of CEI to not only halogens but also biologically-relevant functional groups such as the hydroxyl, -OH, group, which paves the path towards future investigations of larger, biologically-relevant molecules by CEI coupled with covariance analysis.

In recent years, covariance mapping of CEI studies has led to further developments such as (i) the capability to image photodissociation dynamics in real-space and real-time by time-resolved covariance mapping, and (ii) visualizing the three-dimensional fragmentation dynamics of multiply-charged Coulomb exploding molecule by 3D covariance mapping. As a typical example of such recent development, Fig. 13a shows the time-resolved $Cov(I^+, CH_2Cl^+)$ covariance maps resulting from the UV-photodissociation of the C-I bond in $CH_2ICl$ that was subsequently probed by CEI with an 800-nm laser pulse.[80] At negative pump-probe delays (i.e. when the probe pulse arrives before the pump pulse), the intact molecule Coulomb explodes to generate a high energy $CH_2Cl^+$ particle that is emitted back-to-back to the $I^+$ ion. While this high-energy signature is also always present at positive pump-probe delays (i.e. when the pump pulse arrives before the probe pulse). We observed an additional signal in the covariance map with a lower velocity which moves towards the centre of the recoil-frame map. Thus, time-resolved recoil-frame covariance imaging allows to track molecular structure changes in real-space.

The added potential of 3D covariance imaging[83,102] is demonstrated in Fig. 13b that displays the 3D recoil emission direction of $I^+$ ions relative to the $CF_3^+$ reference following the electron impact ionization of $CF_3I$.[83] The 3D recoil velocities show that the two-body dissociation of $CF_3I$ into the $I^+$ and $CF_3^+$ ion fragments are emitted back-to-back to one another in 3D space. Future applications of 3D covariance mapping could potentially reveal out-of-plane velocity components in many-body dissociation processes (i.e. in larger, more complex molecules) which would otherwise not be observable with 2D covariance mapping.

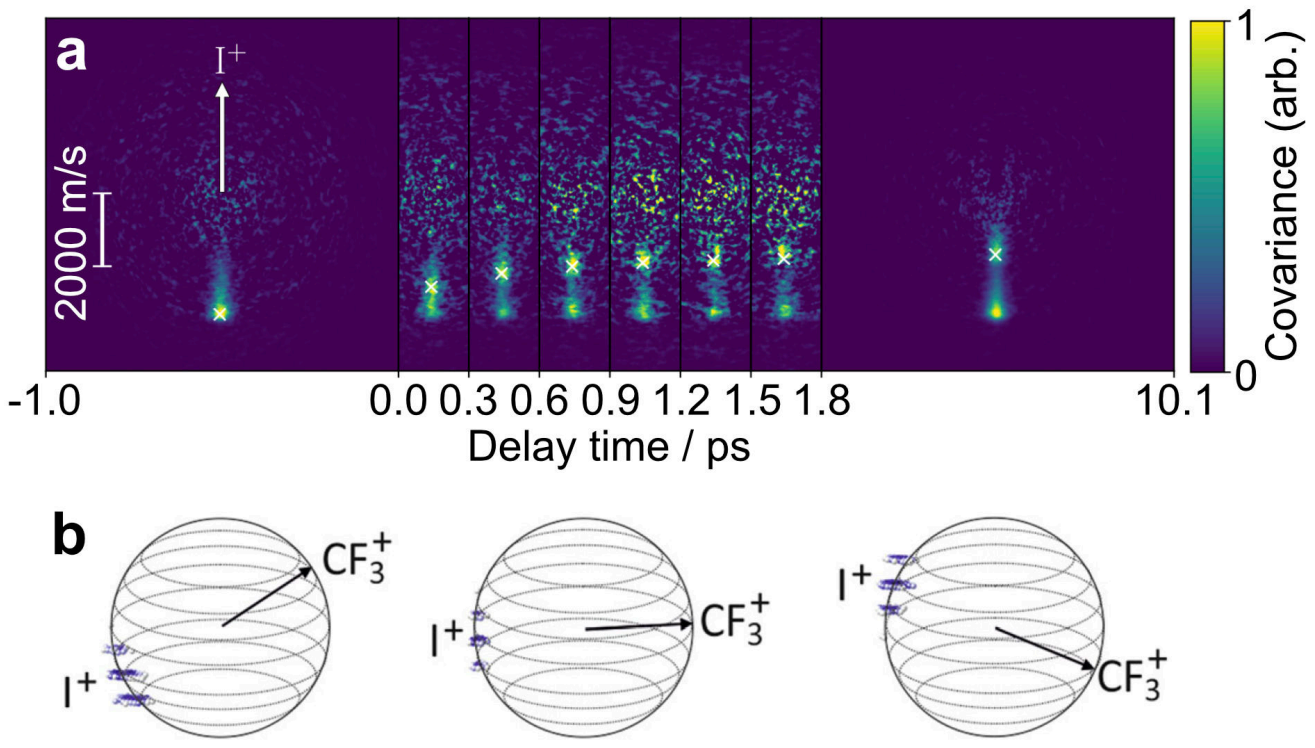

**Figure 13.** Time-resolved and 3D covariance imaging. (a) Measured time-resolved recoil-frame $Cov(I^+, CH_2Cl^+)$ covariance maps following the UV-photodissociation and subsequent CEI-probe of $CH_2ICl$ molecules. **(b)** Three-dimensional (3D) covariance mapping of two-body fragmentation following the electron-impact ionization of $CF_3I$. Panels (a) and (b) adapted from [80] and [83], respectively.

### *3.4 CEI studies with free-electron lasers (FELs)*

Typical In all the measurements presented so far, Coulomb explosion was induced by intense, femtosecond laser pulses with a wavelength in the visible and near-infrared (NIR) spectral range. Coulomb explosions of a molecule by an intense, femtosecond XUV and X-ray pulses from facility-based free-electron lasers (FELs) have also been intensively investigated.[76,77,79,81,82,85–87,92,218] As previously mentioned, with respect to strong field ionization by an intense, NIR laser pulse, ionization by high energy photons proceed via single-photon absorption, or a multiphoton (direct or sequential) absorption process, both can be accompanied by Auger emission. Fig. 14 shows the momentum distributions of $F^+$, $I^{3+}$, $I^{2+}$ and $I^+$ ions following the CEI of 2,6-difluoroiodobenzene (2,6-DFIB) at 800-nm and 11.6-nm for various alignment and orientation distributions of the molecule.[76] In general, the Coulomb explosion of 2,6-DFIB induced by 800-nm and 11.6-nm ionization were observed to be similar despite the very different ways in which the ionization occurs.[76] We note nevertheless that core-shell ionization at high photon energy generally leads to the production of fragments with higher charges. XUV and X-ray CEI has been used to track in real-time the photo-induced dynamics of various molecules,[77,79] such as the charge transfer, photodissociation, and ring-opening dynamics. Given the relatively low repetition rate of most FELs (i.e. << 1 kHz), coincidence measurements is very challenging and instead, covariance analysis has been used.[214,215] We note that the development of high repetition rate X-ray free-electron lasers such as the Eu-XFEL in Hamburg, Germany and LCLS2 in Stanford, USA, provide opportunities to perform high repetition rate coincidence measurements.

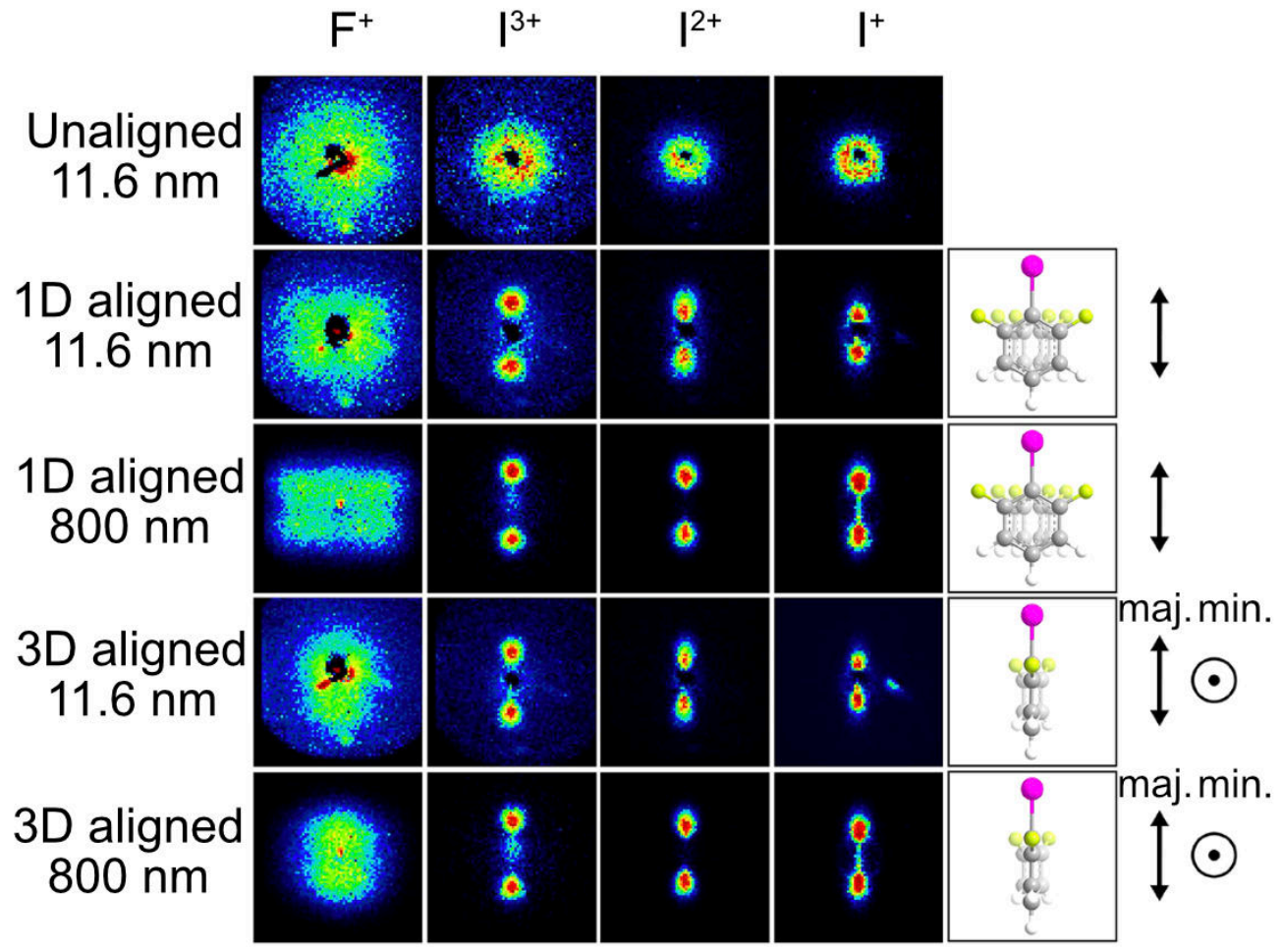

**Figure 14.** CEI by inner-shell core and valence shell ionization. VMI images of $F^+$, $I^{3+}$, $I^{2+}$ and $I^+$ ions generated by the ionization and Coulomb explosion of 2,6-difluoroiodobenzene (2,6-DFIB) using 11.6 nm or 800 nm femtosecond pulses measured with the PImMS1 camera. The corresponding VMI images measured from unaligned, 1D-aligned and 3D-aligned 2,6-DFIB molecules are presented. In 3D alignment, the direction of the major and minor polarization axes of the Nd:YAG adiabatic alignment elliptically polarized pulse are indicated. Figure adapted from [76].

### *3.5 Future perspectives of CEI*

Typical The ultraintense (>>$10^{14}$ Wcm$^{-2}$) pulse intensities provided by today's X-ray FELs (XFELs) enable the generation of very-high charge states of the molecule (>> 2+) suitable for instantaneous and clean Coulomb explosions. For example, ultraintense (up to $10^{19}$ Wcm$^{-2}$, 10 mJ, 1 µm focal spot) and ultrashort (<20 fs) XFEL pulses can be delivered at the small quantum system (SQS) end station of the European XFEL (Eu-XFEL). Such a high peak intensity CEI probe pulse will generate very-high charge states of the molecular ion (i.e. >> +2 charge) to ensure an instantaneous Coulomb explosion of molecules. For example, Fig. 15 shows the calculated recoil-frame $F^+$-$F^+$ correlation maps generated from the Coulomb explosion of the *trans* and *cis* isomers of 4,4'-difluoroazobenzene (DFAB) for various different average charge states of the DFAB molecular ion, assuming that the charges are distributed randomly over the atoms [69]. In this calculation, the trajectory of each ion fragment is calculated using Newton's equations by solving for the particle's velocities. Fig. 15 shows back-to-back emission of the two $F^+$ ions in $Cov(F^+, F^+)$ map corresponding to the *trans* isomer. Whilst in the *cis* isomer, the second $F^+$ ion is emitted 180° and 110° relative to the first $F^+$ reference ion [69]. Moreover, the sideways $F^+$-$F^+$ emission in *cis*-DFAB is more clearly identifiable with high charge states of above +10. This has significant implications in the context of a time-resolved ultraviolet-pump and CEI-probe measurements of *trans-cis* isomerization in DFAB. Specifically, the calculated results of Fig. 15 demonstrate the importance of achieving high charge states (> +10) to distinguish between a transient *cis*-DFAB and an unexcited *trans*-DFAB in a time-resolved measurement since less than 20% of molecules can be typically optically pumped. Future applications of XFEL-based CEI include performing time-resolved pump-probe measurements with the ultraintense (up to $10^{19}$ Wcm$^{-2}$) XFEL probe pulse to generate very high charge states of the molecular ion. This will enable clean and ultrashort (sub-50fs) CEI measurements of the evolving transient structure in photo-induced chemical reactions.

The CEI measurements reported so far have successfully determined the molecular structure of small molecules typically containing less than 12 atoms with atomic spatial resolution. Moreover, the nuclear dynamics of photodissociation reactions have often been investigated, with the ability to track in real-time changes in the dissociating bond length with Ångstrom spatial and <100-fs temporal resolution. Nevertheless, many opportunities exist in future CEI work on ultrafast nuclear dynamics investigations. For example, the combination of CEI with 3D covariance (or coincidence) imaging in pump-probe experiments to the study of gas-phase samples of biomolecules (e.g. amino acids, carbohydrates etc) will allow the 3D structural dynamics of more complex photochemical reactions other than photodissociation (e.g. photoisomerization, photo-induced ring-opening/ring-closing reactions, photo-induced roaming) to be tracked in real-time. Such experiments can be realized with the recent development of high repetition rate, ultraintense (up to $10^{19}$ Wcm$^{-2}$) X-ray FELs (EuXFEL/LCLS II). The advantage of performing CEI using ultraintense X-rays is that the highly-charged molecular ion (> +10 charge) can be fully fragmented into atomic ions that can be simultaneously measured in coincidence/covariance to yield the 3D molecular structure dynamics in photochemical reactions.

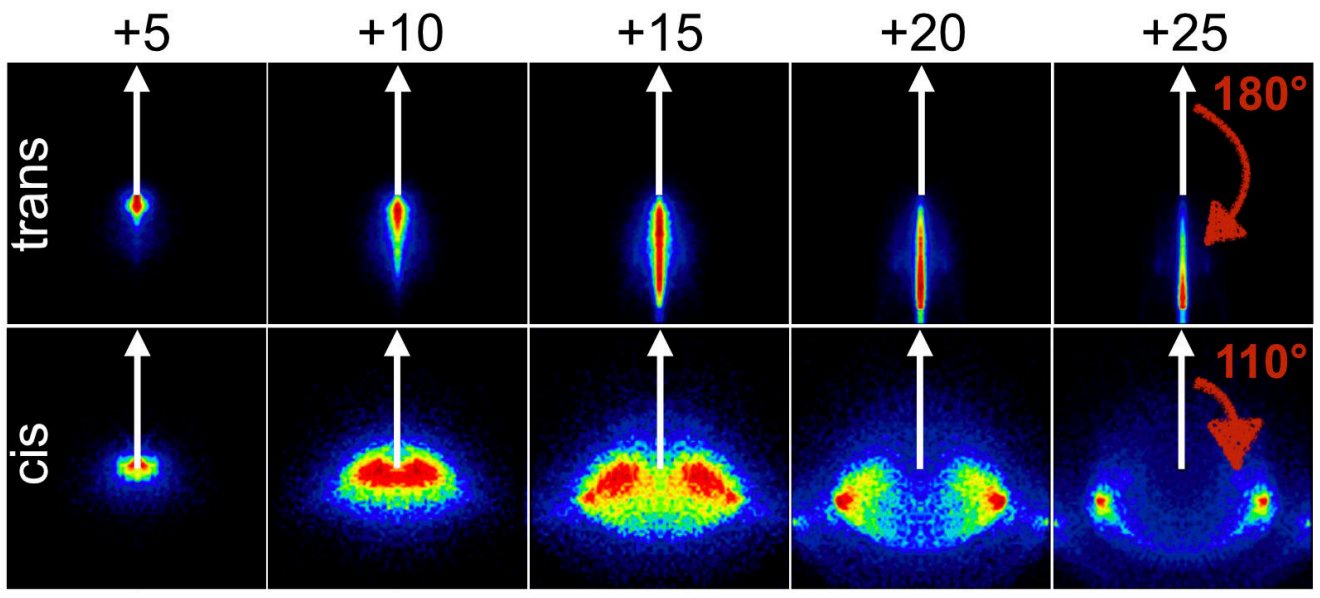

**Figure 15.** Calculated two-fold $F^+$-$F^+$ correlation maps for *trans-* and *cis*-4,4'-difluoroazobenzene (DFAB). The momentum vector of the reference (i.e. first $F^+$) ion is rotated and aligned to the white arrow, whilst the plotted (i.e. second $F^+$) ion is shown in the recoil-frame of the reference ion [69]. The $F^+$-$F^+$ recoil-frame emission angle, $\theta_{F^+F^+}$, is indicated in red. Charge state of DFAB is shown at the top of the figure. Figure adapted from [69].

## 4. Laser-induced electron diffraction

Laser-induced electron diffraction (LIED) [3–7,10,109–127,219] is a strong-field ultrafast electron diffraction (UED) approach that directly retrieves the molecular structure with sub-Ångstrom and femtosecond spatio-temporal resolution. LIED does not rely on the Coulomb explosion of a multiply-charged molecule, and instead uses elastic electron scattering to determine the geometric structure of the singly-charged parent ion. LIED is based on the laser-driven electron recollision process, with Fig. 16 showing the quantum mechanical nature of the recollision process that follows the three-step semi-classical model. Firstly, the field-free electron density distribution in an argon $p$ atomic orbital (see $t = -50$ fs) is exposed to a strong laser field, causing strong field ionization and the subsequent emission of an electron wave packet (EWP; see EWP's wave fronts indicated by the blue area at $t = 0$ fs). The EWP is emitted at portions of the laser's electric field with an intensity equal to or greater than the ionization potential of the argon atom, leading to an appreciable ionization probability (see red shaded at $t = 0$ fs). Secondly, a portion of the emitted EWP is accelerated and returned back to the original electron density by the oscillating electric field. Thirdly, at three quarters of an optical cycle after ionization with a 2.0 μm laser field ($t = +5$ fs), the returning EWP elastically rescatters against the parent ion, generating a rescattered EWP with wave fronts containing interference patterns (see blue area at $t = +5$ fs). Structural information of the target is embedded within the momentum distribution of the rescattered electron which is subsequently detected with a particle spectrometer (e.g. VMI or COLTRIMS REMI). The development of longer, mid-infrared laser sources have enabled the generation of electrons with relatively higher kinetic energies (i.e. $10^2$ – $10^3$ eV order of magnitude). The relatively higher kinetic energies are sufficient to penetrate past the valence electrons, such that rescattering occurs against the core electrons under the first Born approximation. Thus, rescattering occurs close to the nucleus of the atom, and combining this with the sub-Å de Broglie wavelength of the rescattering electron, the atomic positions within the parent molecule can be determined with sub-Å spatial resolution at the time of rescattering, $t_r$. Since ionization and recollision typically occurs within a half-cycle of the oscillating laser pulse, a time resolution on the attosecond-to-femtosecond timescale is achievable, which makes the LIED technique a very attractive structural probe of ultrafast nuclear dynamics in molecules.

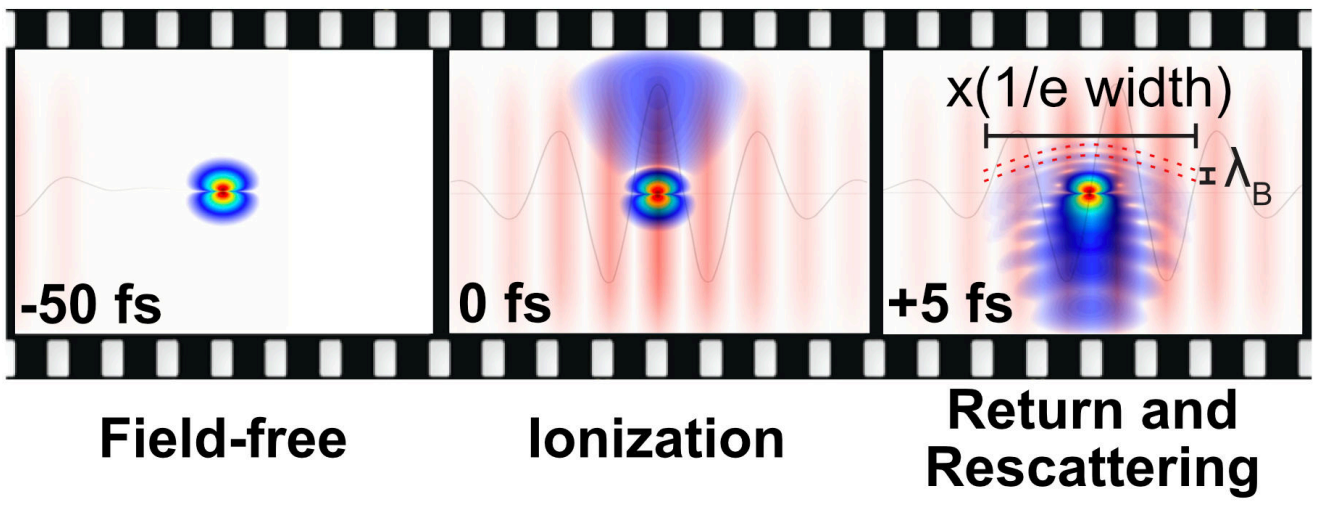

**Figure 16.** Quantum mechanical picture of LIED. A field-free electron density from an Argon $p$ orbital ($t = -50$ fs) is exposed to a strong 2.0 μm laser field which leads to the emission of an electron wave packet (EWP; $t = 0$ fs). The emitted EWP is accelerated and returned by the oscillating electric field, leading to the interference of the EWP and the target ion ($t = +5$ fs). The lateral spread (1/e width) and the de Broglie wavelength of the returning EWP at the instance of rescattering is indicated by $x$ and $\lambda_B$, respectively. Figure adapted from [3].

### *4.1 Elastic electron (re)scattering*

Elastic electron (re)scattering can accurately retrieve the molecular structure only if an electron with a small de Broglie wavelength, $\lambda_B$, is detected within a large enough momentum transfer window, $\Delta q$, in reciprocal space. The wave nature of the scattering electron enables its description with a planar wave $\psi_0 = Ae^{ik_0 z}$ travelling along the $z$ axis with a momentum/wave vector $\vec{k}_0$ and a wavelength $\lambda = 2\pi/k_0$. The momentum transfer, $q$, is defined as the change in the initial momentum vector, $k_0$, of the incoming electron after elastic electron scattering against the target molecule, and is defined as[3,220]

$$q = |\vec{q}| = |\vec{k}_0 - \vec{k}| = 2k_0 \sin\left(\frac{\theta}{2}\right), \quad (17)$$

where $k_0$ and $k$ are the momentum vector before and after scattering, respectively, where $|\vec{k}| = |\vec{k}_0|$ due to electron scattering without energy loss (i.e. elastic scattering), and $\theta$ is the scattering angle, as shown in Fig. 17. The detected total scattering signal, $I_T$, is averaged over all molecular orientations since each molecular orientation possess equal likelihood of existing, with elastic scattering considered to occur locally on each individual atom under the framework of the independent atom model (IAM). However, to retrieve the molecular structure, the total scattering signal is considered in the framework of two-centre scattering between two atoms $i$ and $j$ at an internuclear distance, $\vec{r}_{ij} = \vec{r}_i - \vec{r}_j$, leading to[3,220]

$$I_T(q) = \frac{K^2 I_0}{R^2} \sum_{i=1}^{N} \sum_{j=1}^{N} f_i(q) f_i^*(q) e^{i(\vec{q}\vec{r}_{ij})}, \quad (18)$$

where $f_i(q)$ is the atomic electron scattering amplitude from the partial waves method, given by[3,220]

$$f_i(q) = |f_i(q)| e^{[i\eta_i(q)]}, \quad (19)$$

where $|f_i(q)|$ and $\eta_i(q)$ are the absolute value and phase of the scattering amplitude, respectively, at atom $i$. In general, the total measured interference signal, $I_T$, is composed of the molecular interference signal of interest, $I_M$, a background signal arising from the incoherent sum of atomic scatterings, $I_A$, as given by[3,220]

$$I_T(q) = I_M(q) + I_A(q) = \left( \frac{K^2 I_0}{R^2} \sum_{\substack{i=1 \\ i\neq j}}^{N} \sum_{\substack{j=1 \\ i\neq j}}^{N} |f_i(q)| \, |f_j(q)| cos[\eta_i(q) - \eta_j(q)] \frac{\sin(qr_{ij})}{qr_{ij}} \right. \quad (20)$$

$$+\left(\frac{K^2 I_0}{R^2}\sum_{i=1}^{N}|f_i(q)^2|\right).$$

The physical significance of the $I_M$ and $I_A$ signals are explained using analogies between Young's double slit experiment and elastic electron scattering against a diatomic molecule (grey structure in Fig. 17). Here, the molecular signal $I_M$ arises from the constructive and destructive interference between the resultant waves after scattering against a diatomic molecule (analogous to the double slit in Young's experiment), and is dependent on the internuclear distance between the two atomic scattering centres (analogous to the slit distance in Young's experiment). The background signal $I_A$ is the incoherent sum of electron scattering against only a single atom (analogous to scattering of only one slit in Young's experiment) which contains no structural information. The background atomic $I_A$ signal in fact dominates the total measured $I_T$ signal. The molecular contrast factor (MCF) given by[3,220]

$$MCF = \frac{I_T - I_A}{I_A} = \frac{I_M}{I_A}, \quad (21)$$

$$MCF = \frac{1}{\sum_{i=1}^{N}|f_i(q)^2|}\sum_{\substack{i=1\\ i\neq j}}^{N}\sum_{\substack{j=1\\ i\neq j}}^{N}|f_i(q)|\,|f_j(q)|\cos[\eta_i(q) - \eta_j(q)]\frac{\sin(qr_{ij})}{qr_{ij}}, \quad (22)$$

accounts for the background $I_A$ signal and it also compensates for the significant reduction in scattering intensity with increasing $q$ by contrasting the molecular $I_M$ signal to atomic $I_A$ signal. Structural information from the interference signal can be obtained by Fourier transforming the MCF to generate the radial distribution function given by[3,220]

$$g(r_{ij}) = \int_0^{q_{max}} MCF(q)\, e^{-\beta q^2}\sin(qr_{ij})\, dq, \quad (23)$$

where $\beta$ is a dampening factor that ensures the transform of the MCF between the limits $[q_{max}, \infty]$ is dampened to zero, and $q_{max}$ is the maximum momentum transfer given by[3,220]

$$q_{max} = \frac{2\pi}{\Delta r_{ij}}, \quad (24)$$

where $\Delta r_{ij}$ is the spatial resolution.

To highlight the importance of a large momentum transfer window $\Delta q$, let's consider the elastic electron scattering of carbon dioxide ($CO_2$), as shown in Fig. 18. Firstly, it should be noted that each two-centre scatterer in the molecule possesses a sinusoidally oscillating MCF interference signal, as shown in Fig. 18a and Eqn. (22). The measured MCF signal is equivalent to the sum of the MCF distributions arising from each two-centre scatterer (black distribution). The calculated MCF is shown for a momentum transfer $q$ of up to 30 Å$^{-1}$ (Fig. 18a) together with its corresponding radial distribution (Fig. 18b).[220] Here, short internuclear distances (e.g. C=O bond length of ~1.1 Å$^{-1}$ in Fig. 18b) in radial space possess large wavelength sinusoidal oscillatory interference signals in reciprocal space (see black distribution in Fig. 18a), and thus, require a large momentum transfer window (i.e. $\Delta q = 6\ \text{Å}^{-1}$; e.g. at least $0 - 6\ \text{Å}^{-1}$ range for C=O bond in $CO_2$) to accurately retrieve the molecular structure. For larger internuclear distances (e.g. O-O distance in $CO_2$), a smaller momentum transfer window (i.e. $\Delta q = 3\ \text{Å}^{-1}$; e.g. $0 - 3\ \text{Å}^{-1}$ range for O-O distance in $CO_2$) will suffice. In general, bond lengths typically span 1-3 Å in molecules, and so a maximum momentum transfer and momentum transfer window of $q_{\text{max}} > 6\ \text{Å}^{-1}$ and a $\Delta q > 5\ \text{Å}^{-1}$, respectively, together with a sub-Å de Broglie wavelength are required for accurate structural retrieval.

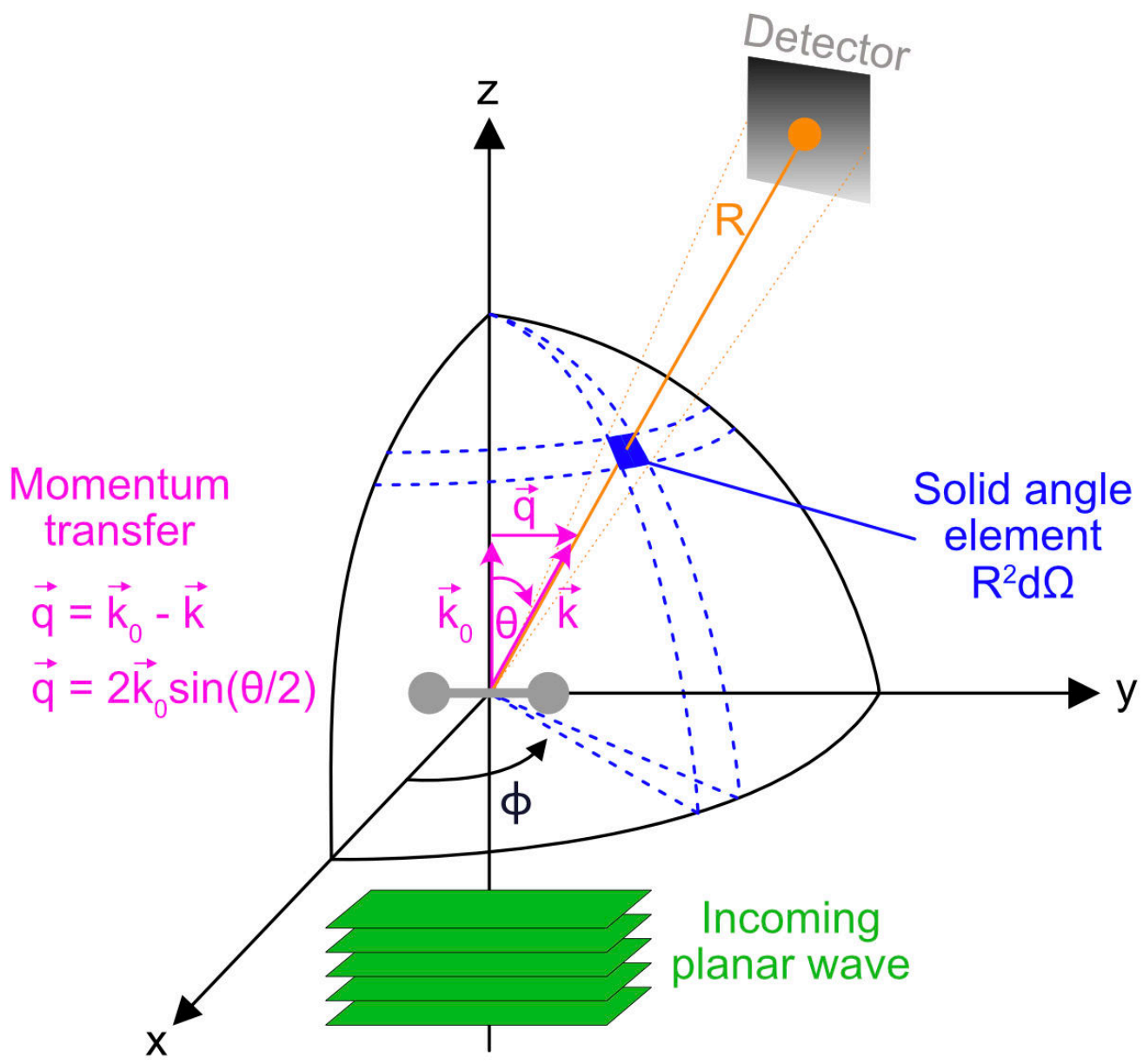

**Figure 17.** Schematic of elastic electron scattering against a diatomic molecule. An incoming planar wave (pink rectangular planes) representing an electron with an initial momentum $\vec{k}_0$ scatters against a generic diatomic molecule. The incoming scattering electron is represented as a plane waves (green rectangular planes) with an initial momentum $\vec{k}_0$ that elastically scatters against a diatomic molecule (grey structure at origin). The electron is scattered into the scattering, θ, and azimuthal, ϕ, angles with a scattered momentum $\vec{k}$, leading to a momentum transfer, $\vec{q} = \vec{k}_0 - \vec{k} = 2k_0\sin(\theta/2)$, (see pink) between the electron and target molecule. The scattered electron is measured within a solid angle element $R^2 d\Omega$ (blue filled square) on a detector (grey rectangle) of distance $R$ away from the scattering point through a solid angle $\Omega$. The number of measured electrons that are scattered into a specific solid angle element $R^2 d\Omega$ is related to the probability of a hard collision occurring, called the scattering cross-section, $\sigma$. The cross-section is typically studied as a function of the electron's angle and energy in electron diffraction measurements which gives the doubly-differential cross-section (DCS), $d^2\sigma/d\Omega^2$. The total elastic cross-section, $\sigma_{2D}$, is obtained by integrating the DCS over all solid angles. Figure adapted from [3].

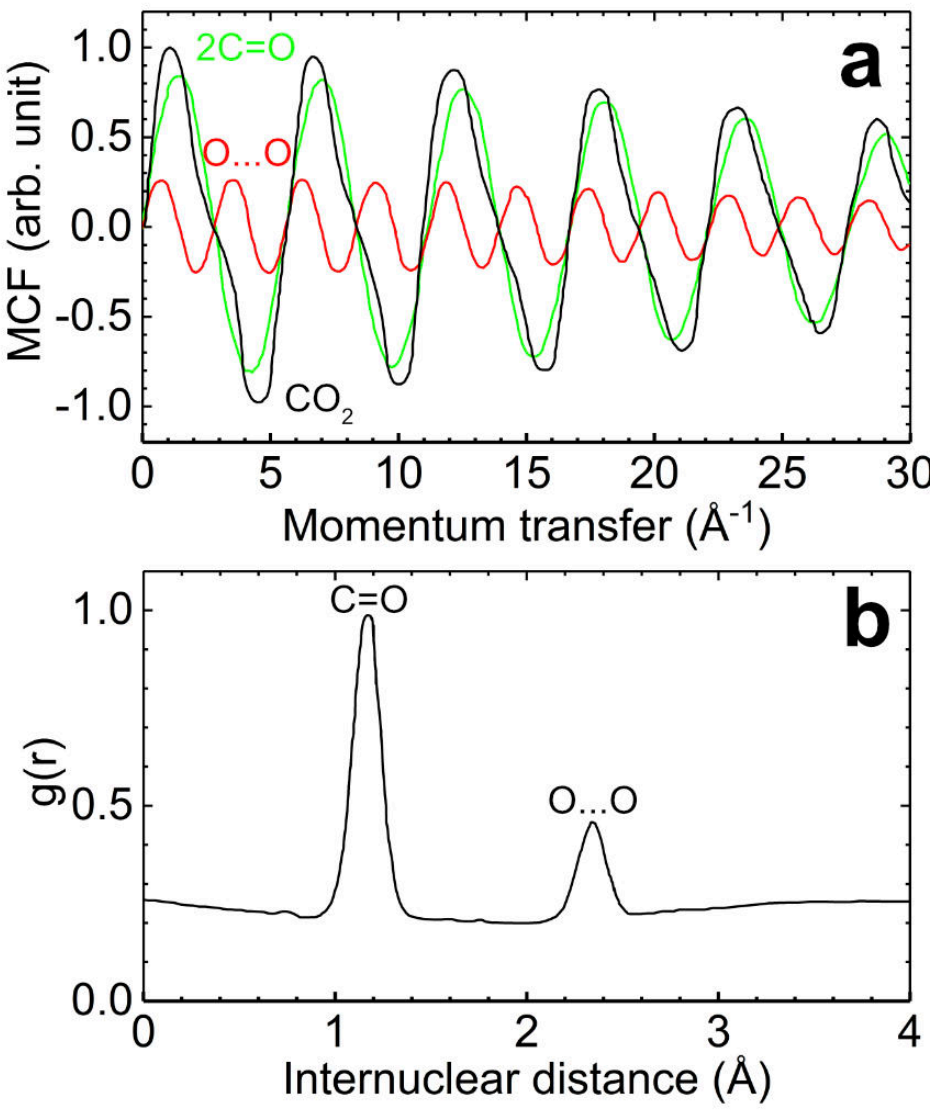

**Figure 18.** Interference and radial distributions from the elastic electron scattering of $CO_2$. (a) The molecular contrast factor (MCF) interference signal as a function of momentum transfer, q, for the C-O (green) and O-O (red) internuclear distances together with their sum (black). (b) Radial distribution, $g(r)$, as a function of internuclear distance. The internuclear distance of C-O and O-O are shown at 1.16 Å and 2.32 Å. Figure adapted from [220].

### *4.2 Mid-infrared LIED*

In LIED, a large momentum transfer window and sub-Å de Broglie wavelength can be achieved by using long driver wavelength laser sources. Fig. 19a shows that at 3.0 μm, electrons can be generated with sub-Å de Broglie wavelength and high kinetic energies spanning hundreds of eV. Fig. 20a shows that 100 eV electrons can achieve a momentum transfer of approximately between 2 – 10 Å for sufficient for accurate structural retrieval. Generating such high kinetic energies is due to the $\lambda^2$ scaling of the ponderomotive energy, $U_p$, which arises from the larger distance and longer acceleration time of the free electron in the oscillating electric field of a longer (e.g. mid-infrared, MIR) wavelengths laser pulse. Standard femtosecond laser systems employ Ti:Sapphire technology which have a central wavelength in the near-infrared (NIR) regime of $\sim$0.8 μm, leading to typical electron kinetic energies of tens of eV and a $\Delta q \approx 4 - 5\ \text{Å}^{-1}$. Thus, near-infrared sources are therefore insufficient for LIED structural retrieval, requiring longer, MIR driver sources. Performing strong-field physics (SFP) measurements such as LIED in the MIR have additional benefits. Fig. 19c shows that at longer wavelengths (e.g. 3.0 μm), SFP measurements can be performed deeper into the quasistatic (tunnel ionization) regime (i.e. $\gamma \ll 1$), leading to highly-energetic, classical electron trajectories and enabling the use of classical recollision models in describing SFP phenomena such as LIED. For example, the quantitative rescattering (QRS) model [7,221,222] can calculate and map electron trajectories to measured experimental features, which provides the propagation times and energy of the returning electrons at the instance of rescattering. This enables the study of changes in intensity of the electron signal at different kinetic energies arising from the molecular interference signal, $I_M$, which is dependent on the molecular structure (i.e. $I_M$ depends on the internuclear distance of two-centre scatterings). The quasistatic regime (i.e. $\gamma \ll 1$) can be reached by increasing the peak intensity of the laser field. However, this can lead to significant ionization and fragmentation of the molecular ion, depleting the number of ground-state neutral molecules readily available to perform LIED measurements with. An alternative is to employ longer driver wavelengths which minimize the fraction of molecules ionized prior to the LIED process whilst operating deep in the quasistatic regime [205]. For example, the ionization fraction of acetylene (with an ionization potential of 12 eV) is 0.3% at 3.0 µm as compared to 100% at 0.8 µm. Performing LIED measurements in the MIR regime has several challenges. The main challenge is the significantly lower rescattering cross-section, $\sigma_r$, of electrons generated in the MIR due to the $\lambda^{-4}$ scaling of the rescattering process. This challenge is circumvented by employing high repetition rate laser sources in the MIR regime. For example, in a typical SFP measurement, a similar acquisition time can be achieved at 3.0 µm with a repetition rate of 160 kHz as compared to a measurement at 0.8 µm with a repetition rate of 1 kHz [205].

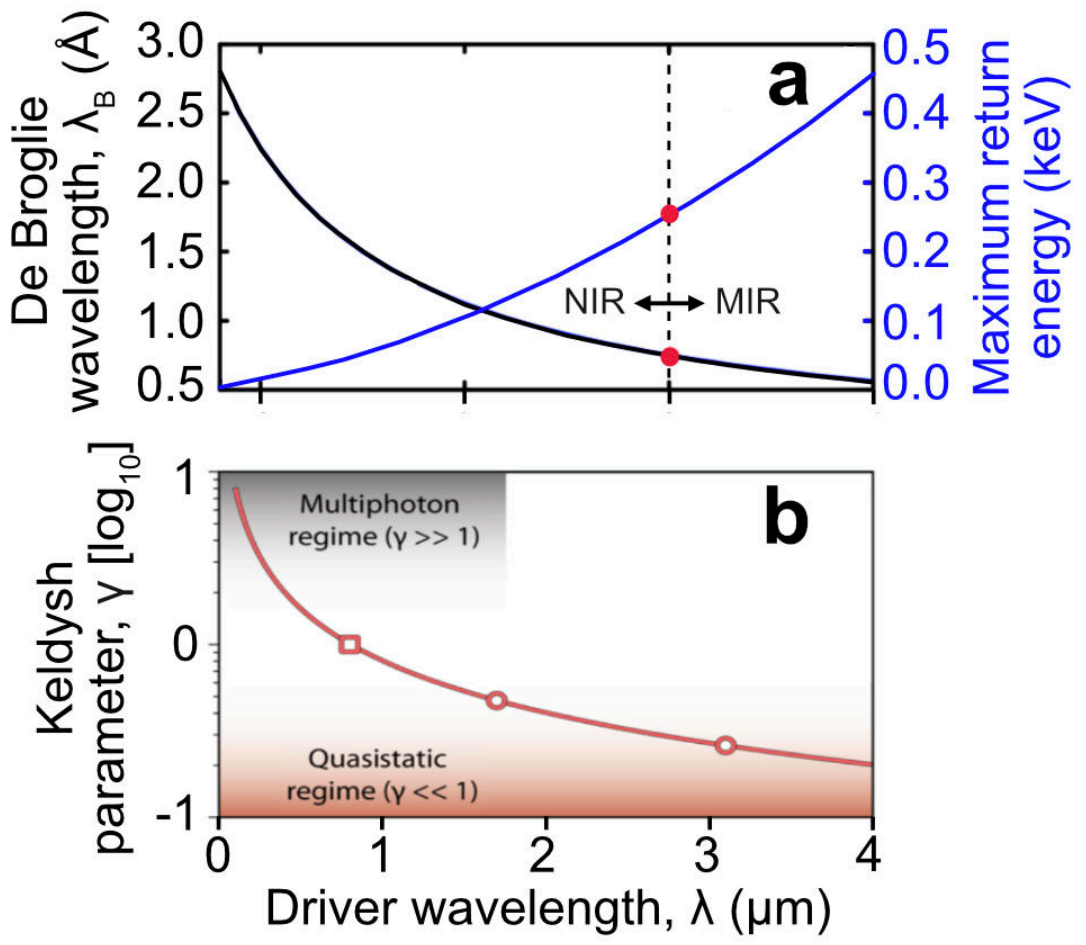

**Figure 19.** Benefits of performing LIED in the MIR regime. (a) The de Broglie wavelength, $\lambda_B$, (black) and maximum return energy (blue) of the rescattering LIED electron as a function of the driver laser wavelength, $\lambda$. The transition from the near-infrared (NIR) and mid-infrared (MIR) regimes is indicated. (b) The Keldysh parameter, $\gamma$, as a function of the driver laser wavelength, $\lambda$. Panel (a) adapted from [223], and panel (b) adapted from [205].

### *4.3 LIED and UED*

LIED can extract field-free DCSs from a field-dressed measurement which are comparable to those measured with field-free ultrafast electron diffraction (UED). Both LIED and UED achieve hard collisions with a momentum transfer of greater than 2 Å$^{-1}$ across a momentum transfer range of approximately 1 - 20 Å$^{-1}$ (see Fig. 20a). However, both methods employ different approaches in achieving high momentum transfers, $q = 2k_0 \sin(\theta/2)$. Fig. 20a shows the typical scattering angle ranges that UED $(0 - 10°)$ and LIED $(40 - 180°)$ operate at. Fig. 20b shows that at low scattering angles $(\theta < 10°)$, comparable scattering amplitudes of elastic electron scattering on carbon and hydrogen atoms are achievable with both UED at 20 keV and LIED at 50 eV, but that the scattering amplitudes for UED quickly decreases at $\theta > 10°$. This is inherently related to the head-on forward scattering geometry and the high kinetic energies employed in UED. In LIED, additional scattering information can be obtained because of the appreciable scattering amplitudes present at a wide range of scattering angles, $\theta$, as shown in Fig. 20b. For example, at $\theta = 45°$, the scattering amplitude from LIED is approximately $10^7$ times larger than that of UED. Thus, LIED is sensitive to the scattering signal as a function of both the kinetic energy and scattering angle of the rescattering LIED electron, enabling the extraction of doubly-differential scattering cross-sections. It should be noted that elastic electron scattering is not only dependent on the scattering energy or angle but also on the identity of the atom that scattering occurs against. For example, hydrogen has a lower scattering amplitude than carbon as demonstrated by the less than unity of the scattering cross-section ratio of carbon to hydrogen, $\sigma_H/\sigma_C$, as shown in Fig. 20c. In fact, LIED has a larger dynamic range in relation to comparable electron scattering amplitudes of both light and heavy atoms in a molecule than in UED because of the larger $\sigma_H/\sigma_C$ value possible with LIED. This avoids replacing hydrogen atoms with heavier "tag" atoms such as halogen atoms, enabling the study of hydrogen atoms in molecules and their pivotal role in biological and chemical processes. It should be noted that UED measurements typically employs orders of magnitudes higher number of electrons per pulse (e.g. $10^4$ - $10^6$ electrons/pulse) than in LIED (one-to-few electrons/pulse), achieving comparable (or higher) electron scattering signal levels to LIED on a pulse-to-pulse basis.

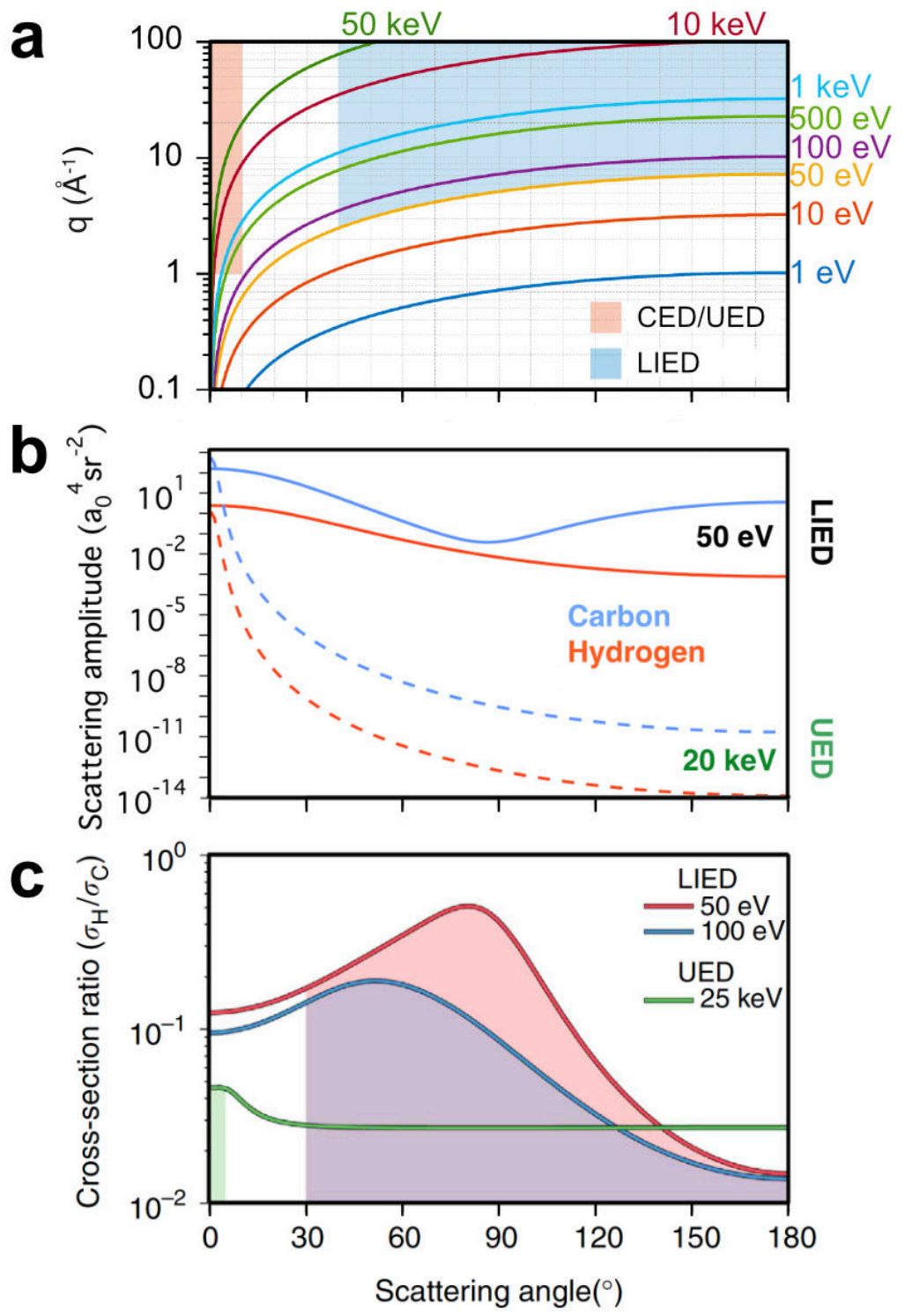

**Figure 20.** Energy-dependence of elastic electron scattering. (a) Momentum transfer, $q$, as a function of scattering angle for kinetic energy ranges of the scattering electron in LIED and CED/UED. The typical regime of operation for LIED and CED/UED is indicated by blue and orange shaded areas, respectively. (b) Scattering amplitude of carbon (blue) and hydrogen (red) atoms as a function of scattering angle for two typical kinetic energies of LIED (solid) and UED (dashed). **(c)** Ratio of scattering cross-sections of hydrogen to carbon atoms as a function of scattering angle for LIED and UED. Panels (a) and (b) are adapted from [3], and panel (c) is adapted from [219].

### *4.4 Tools and data analysis procedures to perform MIR LIED*

Two key tools are required to perform LIED in the MIR regime. Firstly, a high repetition rate (≥ 100 kHz), high average power (> 10 W), femtosecond (≤ 100 fs) MIR (≥ 3.0 µm) laser source [157,158,224–226] will provide intense (> $10^{13}$ Wcm$^{-2}$) pulses to perform the LIED process. Femtosecond laser systems with high average power at high repetition rate typically employ fiber-based technology due to the superior stability and capability (to Ti:Sapphire technology) in managing thermal effects at high average power. Secondly, an electron-ion spectrometer is required that can simultaneously detect the momentum distribution of electrons and ions, ideally under coincidence conditions. The COLTRIMS reaction microscope (REMI) and VMI spectrometers can be employed in LIED measurements, details of which are given in Section 2.4. It should be noted that the COLTRIMS REMI is preferred for LIED measurements since it can directly measure the 3D momentum distributions of both electrons and ions under full kinematic coincidence conditions (i.e. <0.1 molecule fragment per laser shot) with sub-10 meV momentum resolution, ensuring that the generated particles originated from the fragmentation of a single, isolated molecule. Performing coincidence measurements enables the LIED electron signal originating from the parent ion to be isolated from the total electron signal which contains background signal from other, unwanted processes (e.g. Coulomb explosion and fragmentation of the molecular ion).

The remainder of this section will present LIED data measured with a COLTRIMS REMI unless otherwise stated, and the typical data analysis steps employed. Fig. 21a shows the typical time-of-flight (ToF) spectrum following the strong-field ionization of $H_2O$ gaseous molecules integrated over ten billion laser shots. The dominant ToF peak corresponds to the $H_2O^+$ molecular ion, with appreciable background contributions from other ions generated from the Coulomb explosion of $H_2O^{2+}$. Figs. 21b and 21c show the typical data analysis employed in LIED and is based on LIED measurements of $Xe^+$ ions. Fig. 21b shows the measured 3D momentum distribution of electrons detected in coincidence with $Xe^+$ ions in cartesian coordinates. Due to cylindrical symmetry of the LIED measurement, a cartesian-to-cylindrical coordinate transformation is employed, and the signal is integrated along the azimuthal angle $\phi$ (see Fig. 21b). The reduction of the measured 3D momentum data to a 2D distribution simplifies the LIED analysis and enables comparison to theory. In a final step, the correct solid angle $\Omega$ is ensured by applying a Jacobian, $1\backslash|p_\perp|$, where $p_\perp$ is the electron momenta emitted perpendicular to the laser polarization direction. In Fig. 21c, the field-free elastic DCS is extracted by considering the influence of the laser's vector potential, $A_\mathrm{r}(t)$, "kick" on the returning electron with a return momentum, $k_\mathrm{r}$, at the time of rescattering, $t$. The measured rescattered momentum, $k_{resc}$, is in fact the sum of the two momentum components as given by $k_\mathrm{resc} = A_\mathrm{r}(t) + k_\mathrm{r}$. The electron signal is integrated along the circumference of a circle with a radius $k_\mathrm{r}$ that has its origin shifted along the longitudinal momentum, $p_\parallel$, (i.e. the electron momenta emitted parallel to the laser polarization direction) by $A_\mathrm{r}(t)$ to account for the laser's vector potential "kick". Fig. 21c shows the schematic of the electron signal integrated at three different return energies of 75 eV (blue circle), 100 eV (red circle) and 125 eV (green circle). Fig. 21d shows the corresponding integrated electron signal as a function of rescattering angle $\theta_\mathrm{r}$. Here, there is a good agreement between LIED measured (circles, squares, triangles) with reference field-free DCSs from the NIST database (solid lines) [227], confirming LIED's capability to retrieve field-free DCSs under the presence of a laser field.

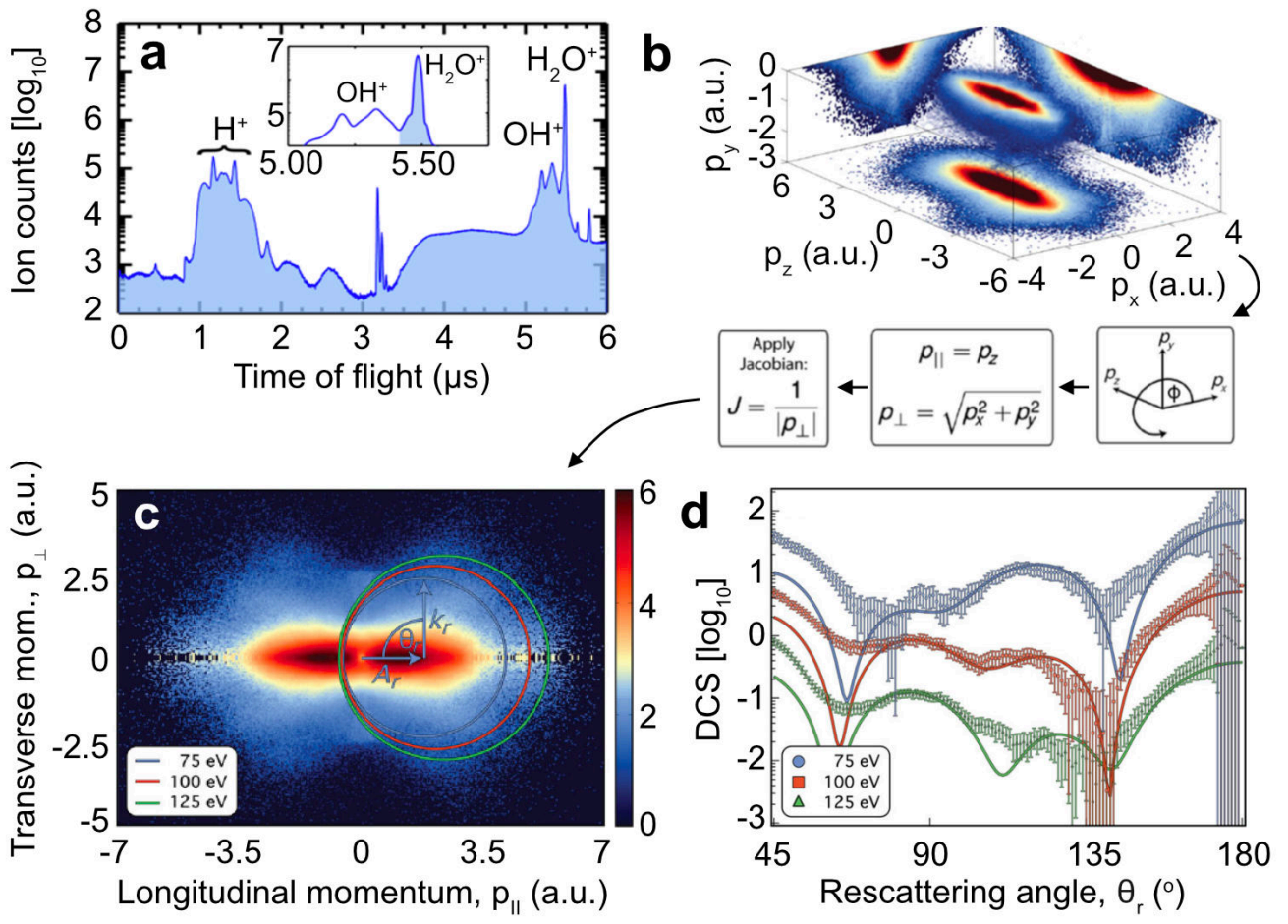

**Figure 21.** Analysis of LIED data measured with a COLTRIMS REMI. (a) Typical time-of-flight (ToF) spectrum measured with a COLTRIMS REMI following the strong-field ionization of gaseous $H_2O$ molecules with a $\sim 1 \times 10^{14}\mathrm{Wcm}^{-2}$, 97 fs (FWHM), 3.2 µm laser pulse [122]. A zoom-in shows the dominant $H_2O^+$ ToF of interest. (b) Three-dimensional (3D) momentum distribution of electrons measured in coincidence with $Xe^+$ ions following the strong-field ionization of Xe by $\sim 7 \times 10^{13}\mathrm{Wcm}^{-2}$, 70 fs (FWHM), 3.1 µm laser pulse. (c) The two-dimensional (2D) momentum distribution generated from the 3D distribution in panel (b) following the analysis procedure outlined briefly in the figure and discussed in detail in the main text. (d) Field-free differential cross-sections (DCSs) extracted from the integration across the circumference of circles in panel (c) as a function of rescattering angle, $\theta_\mathrm{r}$, for three return energies of 75 eV (blue), 100 eV (red) and 125 eV (green). The field-free DCSs (solid lines) from a NIST database are presented as reference data [227]. Panel (a) is adapted from [122] and panels (b)-(d) adapted from [205].

Two main variants of LIED exist, as shown in Fig. 22: (i) quantitative rescattering LIED (QRS-LIED) [3,7,10,116,120,123–125,219,221,222], and (ii) Fourier-transform LIED (FT-LIED) [3,115,122,126] which is also called fixed-angle broadband laser-driven electron scattering (FABLES) [113,123]. In both cases, the background atomic $I_\mathrm{A}$ signal is either calculated or empirically extracted from the measured DCS data in QRS-LIED and FT-LIED/FABLES, respectively.

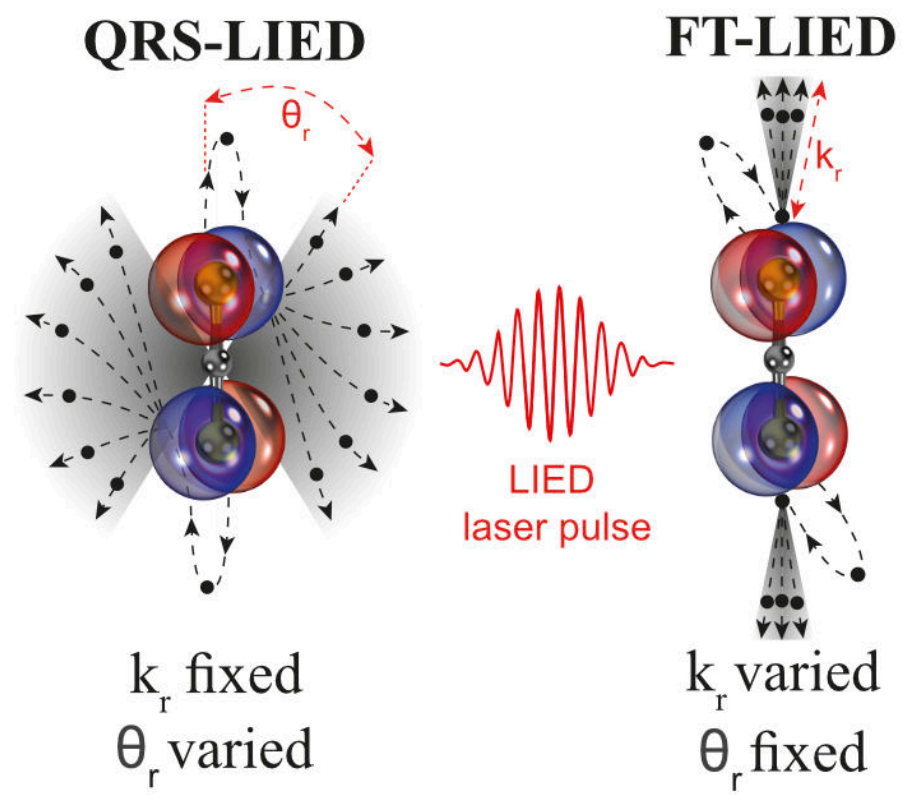

**Figure 22.** Two main variants of LIED. In QRS-LIED (FT-LIED), the returning electron rescatters against the molecular structure at varied (fixed) rescattering angle, $\theta_\mathrm{r}$, and fixed (varied) rescattering momentum, $k_\mathrm{r}$. The molecular structure is retrieved from the Fourier transform of electron signal arising within the rescattering region of interest (grey shaded area). Figure adapted from [223].

### *4.5 QRS-LIED*

In QRS-LIED, [3,7,10,116,120,123–125,219,221,222] elastically rescattered electrons are detected at a wide-range of rescattering angles $\theta_\mathrm{r} = 20 - 140°$ for a fixed return momentum $k_\mathrm{r}$, as shown in Fig. 22. The measured momentum distribution containing the interference signal is then compared to calculated data for known or guessed molecular structures based on the semi-classical quantitative rescattering (QRS) theory [7,221,222] and the independent atomic model (IAM) [7,15,53,54,221,222,228,229]. A chi-squared best fit between the measured and calculated interference signals determines the most likely and dominant molecular structure contributing to the measured signal, as given by

$$\chi^2(P_1, P_2) = \sum_\mathrm{n} [\gamma^\mathrm{e}(k_\mathrm{r}, \theta_\mathrm{n}) - \gamma^\mathrm{t}(k_\mathrm{r}, \theta_\mathrm{n})]^2, \quad (25)$$

where $P_1$ and $P_2$ are the molecular structure parameters in which the chi-squared fitting occurs over, and $\gamma^\mathrm{e}$ and $\gamma^\mathrm{t}$ are the experimentally measured and theoretically calculated molecular contrast factor (MCF), respectively.

Fig. 23 shows the molecular structure retrieval of acetylene ($C_2H_2^+$ [219], $C_2H_2^{2+}$ [10]) and carbon disulfide ($CS_2^+$ [120]) using QRS-LIED, with many other QRS-LIED results also previously reported [111,113,116,117,123–125]. Figs. 23a and 23b show that the retrieved acetylene monocation $C_2H_2^+$ and dication $C_2H_2^{2+}$ molecular structures are significantly different to their equilibrium structures. The $C_2H_2^+$ monocation structure retrieved with LIED (e.g. $R_\mathrm{CH} = 1.05 \pm 0.03$ Å) was found to be in good agreement with its equilibrium structure (e.g. $R_\mathrm{CH} = 1.08$ Å) [230]. Whilst in the case of the $C_2H_2^{2+}$ dication, C-H bond lengths were retrieved ($2.31 \pm 0.15$ Å) that are significantly longer than in the equilibrium structure (1.06 Å) [231], arising from the deprotonation of $C_2H_2^{2+}$ in the presence of the strong 3.1 μm laser field. In fact, other examples of strong-field induced molecular structure modification have also been reported [120,123]. For example, Fig. 23c shows the retrieval of a strongly bent ($\theta_\mathrm{SCS} = 104 \pm 20.2°$) and symmetrically stretched ($R_\mathrm{CS} = 1.86 \pm 0.23$ Å) field-dressed carbonyl disulfide $CS_2^+$ monocation structure [120] that significantly deviates from the linear field-free structure [232]. The presence of a strong laser field enabled a linear-to-bent transition in the molecule on the rising edge of the laser field. Such a transition was assisted by the strong-field assisted Renner-Teller effect [120] where a previously dipole-forbidden transition in the linear geometry became dipole-allowed in the bent geometry. In general, QRS-LIED can retrieve the molecular structure with picometre and femtosecond spatio-temporal resolution, but is limited to small molecules since it requires the calculation of the interference signal and its comparison to measured data using fitting algorithms, which quickly becomes computationally expensive with increasing molecular complexity.

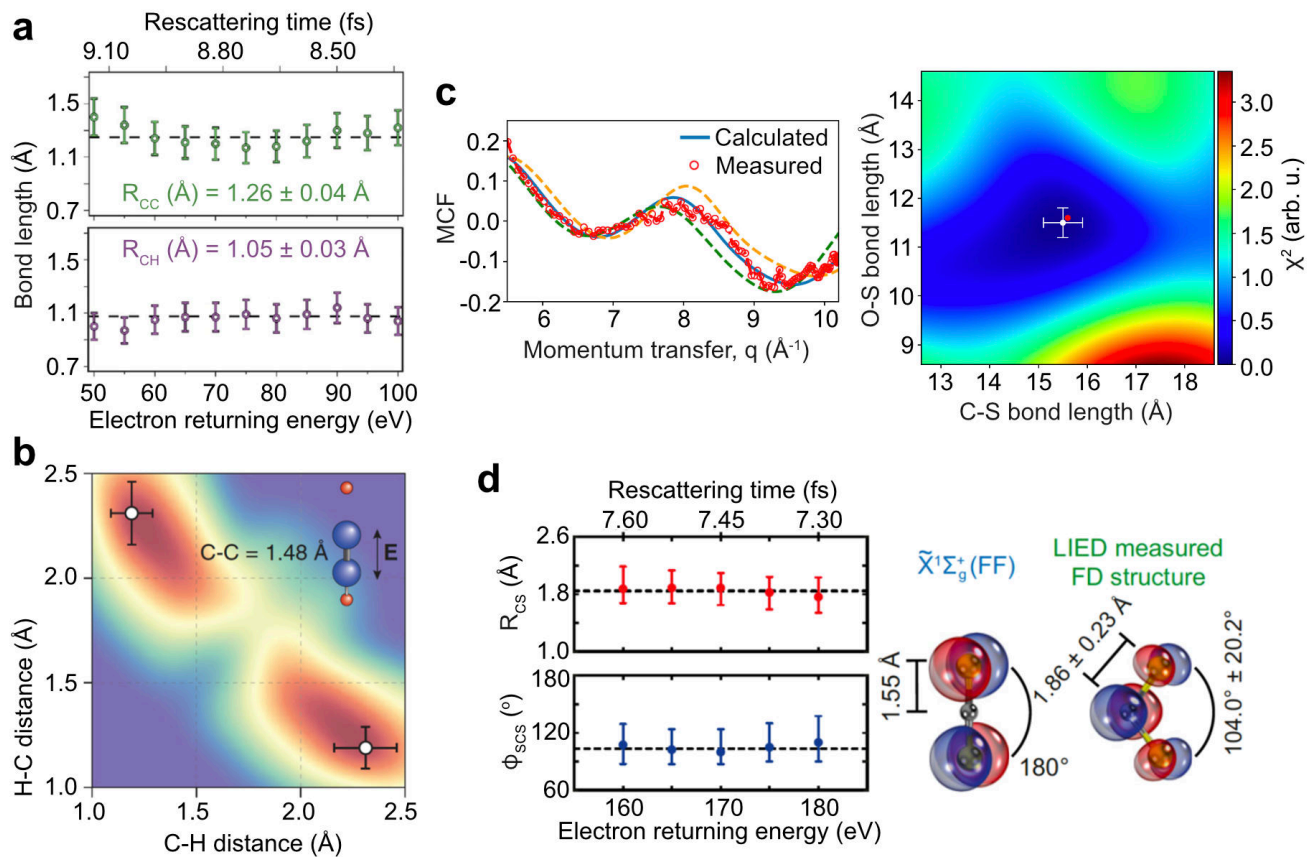

**Figure 23**. Molecular structures retrieved with QRS-LIED. (a) Molecular structure retrieval of the acetylene $C_2H_2^+$ monocation as a function of electron return energy for C-C and C-H internuclear distances, $R_\mathrm{CC}$ and $R_\mathrm{CH}$, respectively. The corresponding equilibrium values are shown as dashed black lines. (b) Two-dimensional $\chi^2$ fitting maps for the acetylene $C_2H_2^{2+}$ dication measured parallel to the polarization axes of the LIED MIR laser field. The C-H internuclear distances corresponding to the best $\chi^2$ fit between the measured and theoretical interference signal is indicated by white circles. (c) Retrieved C-S internuclear distance, $R_\mathrm{CS}$, and S-C-S bond angle, $\phi_\mathrm{SCS}$, as a function of electron return energy from QRS-LIED measurements of field-dressed $CS_2$. The average values of $R_\mathrm{CS}$ and $\phi_\mathrm{SCS}$ are shown as dashed black lines. A schematic of the field-free (FF) linear and LIED measured field-dressed (FD) bent and symmetrically stretched $CS_2$ structure is shown. Panel (a) adapted from [219], panel (b) adapted from [10], and panel (c) adapted from [120].

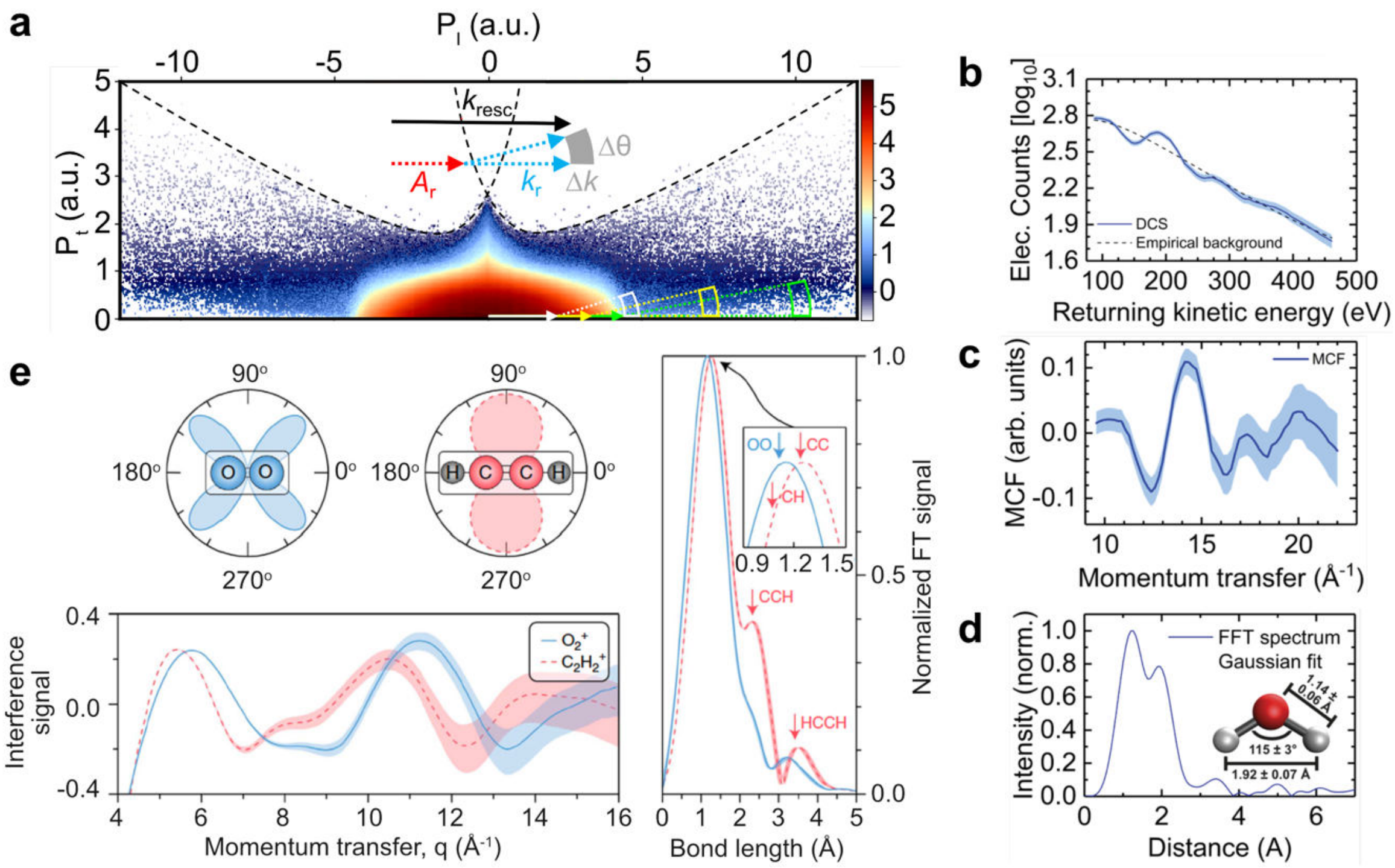

**Figure 24.** FT-LIED retrieval of molecular structure using a COLTRIMS REMI. (a) Longitudinal and transverse momentum (i.e. momenta parallel and perpendicular to the laser polarization axis, respectively) of all electrons emitted from the strong-field ionization of $H_2O$ with a 97 fs, 3.2 μm laser pulse of $1 \times 10^{14}$ Wcm$^{-2}$ peak intensity. The electron signal is integrated within the area given by the block arc of a narrow range in scattering angle and momentum, $\Delta k$ and $\Delta\theta$, respectively. The electron signal is integrated at various measured rescattering momenta, $k_{\mathrm{resc}} = k_{\mathrm{r}} + A_{\mathrm{r}}$, comprised of the return momentum, $k_{\mathrm{r}}$, and the vector potential momentum "kick" of the laser, $A_{\mathrm{r}}$, at the time of rescattering. (b) The integrated electron signal obtained from panel (a) as a function of return kinetic energy of the rescattered electron. The empirical background is given by the dashed black line. (c) The molecular contrast factor (MCF) as a function of momentum transfer. (d) Radial distribution obtained from the Fourier transform of the MCF signal in panel (c). (e) FT-LIED retrieval of $O_2$ and $C_2H_2$ molecular structure independent of the highest occupied molecular orbital (HOMO) symmetry. (Top-left) Calculated ionization probability of strong-field ionized $O_2$ and $C_2H_2$ with $\pi_g$ and $\pi_u$ symmetries, respectively, at various angles relative to the laser polarization direction. (Bottom-left) The MCF distribution as a function of momentum transfer for $O_2^+$ (blue) and $C_2H_2^+$ (red). (Right) Retrieved radial distribution as a function of bond length. Panels (a) and (b)-(d) are adapted from [122] and panel (b) is adapted from [115].

### *4.6 FT-LIED*

In FT-LIED, [3,115,122,126] elastically back-rescattered electrons are measured at a fixed rescattering angle of $\theta_{\mathrm{r}} \approx 180°$ but at various return momentum, $k_{\mathrm{r}}$. This enables a direct image of the object (i.e. the molecular structure) in the near-field to be obtained by Fourier transforming the back-rescattered signal measured in the far-field. Thus, molecular structure retrieval with FT-LIED does not require comparison to calculated interference data or prior knowledge of the structure. The field-free elastic DCS signal is extracted from the measured two-dimensional electron momentum distribution using the following FT-LIED analysis steps, as illustrated in Fig. 24a for an isolated, gaseous water ($H_2O$) molecule [122]. The number of electrons is integrated within a block arc area at various measured rescattering momenta, $k_{\mathrm{resc}}$, along the longitudinal electron momentum (i.e. momenta parallel to the MIR laser polarization axis) given the back-rescattering nature of FT-LIED ($\theta_{\mathrm{r}} \approx 180°$). Here, $k_{\mathrm{resc}}$ is the sum of the return momentum, $k_{\mathrm{r}}$, and vector potential of the laser field, $A_{\mathrm{r}}(t)$, at the time of rescattering, $t$. The integrated electron signal is equivalent to the field-free total DCS, as shown in Fig. 24b, which is composed of the oscillating molecular interference signal, $I_{\mathrm{M}}$, and a background atomic signal, $I_{\mathrm{A}}$, that decreases with increasing momentum transfer, $q$. The background $I_{\mathrm{A}}$ signal is empirically fitted and subsequently used to determine the measured MCF, as shown in Fig. 24c. The radial distribution is obtained by Fourier transforming the MCF interference signal, as shown in Fig. 24d. Here, a slightly bent ($\theta_{\mathrm{HOH}} = 115 \pm 3°$) and stretched ($R_{\mathrm{OH}} = 1.14 \pm 0.06$ Å) $H_2O^+$ molecular structure is retrieved with FT-LIED which is assigned to $H_2O^+$ in the ground electronic state.

In general, a VMI spectrometer can measure FT-LIED/FABLES (and QRS-LIED) data, however, the rescattered electron signal is embedded within the momentum distribution of all detectable electrons, leading to possible contamination or the inability to extract the rescattered electron signal. Fig. 24e demonstrates the significance of filtering the LIED electron signal from all detected electrons through kinematic coincidence measurements in $O_2$ and $C_2H_2$ using a COLTRIMS REMI. Here, clear oscillations in the interference (MCF) signal are observed as a function of momentum transfer, which enhanced by measuring rescattered electrons in coincidence with $O_2^+$ (blue solid) and $C_2H_2^+$ (red dashed) ions of interest. Previous work using a VMI spectrometer under non-coincidence conditions reported the inability to retrieve the molecular structure of $O_2^+$ ions as arising from the orbital symmetry of molecule's highest occupied molecular orbital (HOMO) [113]. However, the COLTRIMS REMI LIED measurement of $O_2^+$ clearly demonstrates that the HOMO orbital symmetry does not impede the retrieval of the $O_2^+$ molecular structure (see radial distribution in Fig. 24e).

### *4.7 Future perspectives of LIED*

Static snapshots of small molecular structures have been successfully retrieved using LIED with picometre and femtosecond spatio-temporal resolution. Two key future research directions for LIED are possible: (i) the retrieval of large, complex molecular structures, and (ii) the implementation of LIED in time-resolved optical-pump LIED-probe measurements of molecular dynamics. The retrieval of large molecular structures (e.g. chiral Fenchone or photoswitches such as stilbene and azobenzene) is challenging with the two main LIED variants of QRS-LIED and FT-LIED. In QRS-LIED, calculating the molecular interference signal for large structures with a sufficiently high spatial resolution becomes computationally expensive. Moreover, a chi-squared fitting algorithm is employed over many different molecular structures to find the structure that gives the best fit between the measured and theoretical interference signal. In FT-LIED, a one-dimensional radial distribution can be directly retrieved by Fourier transform of the molecular interference signal, bypassing the requirement of calculating the interference signal as in QRS-LIED. Bond lengths are extracted from the radial distribution using a multi-peak fitting procedure. For large, complex molecular structures possessing closely spaced bond lengths, it becomes increasingly difficult to identify clear and separable internuclear distance peaks in the radial distribution retrieved by FT-LIED.

New variants of LIED have been developed to surpass the above-mentioned challenges. For example, the zero crossing point variant of LIED (ZCP-LIED)[127] simplifies the LIED molecular structure retrieval process by only considering the critical zero crossing points of the molecular interference signal, $I_{\mathrm{M}}$, which is directly extracted from the laboratory-frame photoelectron spectrum. In the case of carbonyl sulfide (OCS), a minimum of two ZCPs was sufficient to distinguish between the equilibrium linear and measured bent $OCS^+$ geometries. In general, ZCP-LIED avoids the need to fit the whole $I_{\mathrm{M}}$ signal, it bypasses the need to distinguish closely spaced internuclear distances, it avoids converting from different reference frames to retrieve the molecular interference signal, and it can retrieve the molecular structure with lower signal-to-noise data as compared to FT-LIED. A second variant of LIED employs a machine learning (ML) algorithm, called ML-LIED [233], that can predict the three-dimensional (3D) molecular structure. Here, the ML algorithm is trained to find the relationship between the 3D molecular structure and its corresponding two-dimensional differential cross-section (2D-DCS) maps. The training, validation and testing of the ML model is performed with a database of approximately 100,000 structures and their corresponding 2D-DCS maps calculated using the independent atom model. At the heart of the ML model is a neural network that can identify subtle features in the 2D-DCS maps that relate to a common part of the molecular structure. The measured 2D-DCS map is then used as an input of the ML model which predicts the 3D cartesian $(x, y, z)$ coordinates of each atom in the molecule. Here, ML-LIED has successfully predicted the 2D molecular structure of acetylene ($C_2H_2$) and carbon disulfide ($CS_2$) as well as the 3D molecular structure (+)-Fenchone ($C_{10}H_{16}O$; 27 atoms). In the latter case, other LIED variants such as FT-LIED and QRS-LIED were unable to retrieve the molecular structure for such a large and complex molecule as (+)-Fenchone.

Employing LIED in time-resolved optical-pump and LIED-probe measurements will enable the retrieval of multiple high-resolution snapshots of the molecular structure that will generate a picometre and femtosecond "molecular movie" of a photo-induced chemical reaction. Employing the above-mentioned new variants of LIED, in particular ML-LIED, will help identify the contribution of one or more transient molecular structures to the measured time-resolved LIED signal. Furthermore, most photo-induced chemical reactions are initiated in the ultraviolet (UV; 200-400 nm) and vacuum-UV (VUV; 100-200 nm) regime of the electromagnetic spectrum. Optical pulses in the UV-VUV range can be generated using crystal or fiber-based approaches. Typically, a non-tunable 267 nm pulse is generated by the sum frequency mixing of an 800 nm pulse and its 400 nm second harmonic, with a typical pulse duration of >50 fs and pulse energy on the few-to-tens of μJ. Hollow fibers have been employed to generate few-cycle, up to 200 µJ 248 nm pulses [234]. Further developments in the fiber-based approach have led to the generation of tunable UV-VUV pulses with sub-10 fs duration and up to ~10 µJ energy using hollow capillary fibres (HCFs) [235,236] and photonic crystal fibres (PCFs) [237,238]. In general, generating tunable UV-VUV pump and MIR-LIED probe pulses at high repetition rates (>100 kHz) with sufficient pulse energy requires a high-average power laser.

## 5. Concluding remarks

The study of time-resolved gas-phase molecular dynamics can be accomplished with various imaging techniques such as CEI and LIED as well as ultrafast electron and X-ray diffraction. No one single technique is perfectly suited to study all types of molecular structure and dynamics. Instead, employing multiple ultrafast techniques to answer important parts of various scientific questions will reveal the underlying mechanisms and intricacies of many biologically-relevant chemical reactions using the aforementioned ultrafast imaging tools developed in physics. High average power laser sources based on fiber technology enable strong-field CEI and LIED measurements at high repetition rates. In the case of LIED in the MIR regime, the combination of a high repetition rate source with a COLTRIMS and reaction microscope (REMI) specrometer will enable the filtering of the LIED rescattered electron signal can be dominated by background electron signal. The degree of background signal is of course dependent on the type of molecule, the experimental set-up, driver wavelength source, but utilizing the kinetic detection filtering capability of a COLTRIMS and REMI spectrtometer will ensure the LIED signal can be extracted. Institute-based XFEL facilities such as the European XFEL provide ultraintense CEI with XFEL pulses of up to $10^{19}$ Wcm$^{-2}$ peak intensity. Such ultraintense pulses generate very high charge states of the molecular ion, leading to prompt Coulomb explosion and significant fragmentation of the molecule. At such high pulse intensities, atomic fragments are much more likely to be generated than molecular fragments, leading to cleaner and more easily interpretable correlations between ion fragments, and thus simplifying molecular structure identification.

XFEL-based ultrafast X-ray diffraction (UXD) studies have successfully retrieved the static and transient structures of gas-phase molecules. The high flux offered by today's XFELs have enabled such UXD studies. Ultrafast electron diffraction (UED) with tabletop non-relativistic keV and institute-based relativistic MeV set-ups have been employed in time-resolving gas-phase molecular dynamics. Of particular significance, MeV UED measurements at the SLAC facility demonstrated the simultaneous measurement of electronic and nuclear dynamics in photoexcited pyridine molecules by the measurement of inelastically and elastically scattered electrons, respectively.

In general, the combination of results from various experimental techniques combined with state-of-the-art theoretical calculations will further our fundamental understanding of chemical reactivity and chemical bonding.

## Acknowledgements

I would like to thank Arnaud Rouzée, Anja Röder and Luke Maidmount for proofreading this tutorial review.